\documentclass[twocolumn,showpacs,preprintnumbers,amsmath,amssymb,floatfix]{revtex4}

\usepackage[dvips]{graphicx}
\usepackage{array}
\usepackage{dcolumn}
\usepackage{bm}

\newcommand{\lw}[1]{\smash{\lower1.5ex\hbox{#1}}}

\newcommand{\ix}[1]{$#1$}

\newcommand{\benum}{\begin{enumerate}}
\newcommand{\eenum}{\end{enumerate}}
\newcommand{\bit}{\begin{itemize}}
\newcommand{\eit}{\end{itemize}}
\newcommand{\be}{\begin{equation}}
\newcommand{\ee}{\end{equation}}
\newcommand{\bdm}{\begin{displaymath}}
\newcommand{\edm}{\end{displaymath}}
\newcommand{\bqq}{\begin{eqnarray}}
\newcommand{\eqq}{\end{eqnarray}}
\newcommand{\barr}{\begin{array}}
\newcommand{\earr}{\end{array}}

\newcommand{\etapr}{{\eta^{\prime}}}
\newcommand{\etaprime}{{\eta^{\prime}}}

\newcommand{\Ncolor}{N_{\mathrm{color}}}

\newcommand{\Nspin}{N_{\mathrm{spin}}}

\newcommand{\tmin}{t_{\mathrm{min}}}
\newcommand{\tmax}{t_{\mathrm{max}}}

\newcommand{\mev}{\,\mathrm{Me\kern-0.1em V}}
\newcommand{\MeV}{\,\mathrm{Me\kern-0.1em V}}
\newcommand{\gev}{\,\mathrm{Ge\kern-0.1em V}}
\newcommand{\GeV}{\,\mathrm{Ge\kern-0.1em V}}

\newcommand{\mpi}{m_{\pi}}

\newcommand{\Gpi}{G_{\pi}}

\newcommand{\Getapr}{G_{\etapr}}
\newcommand{\Getaprime}{G_{\etapr}}
\newcommand{\Gdisc}{G_{\rm disc}}
\newcommand{\Gconn}{G_{\rm conn}}

\newcommand{\mpiovermrho}{m_{\pi}/m_{\rho}}

\newcommand{\ampi}{am_{\pi}}

\newcommand{\metapr}{m_{\eta^{\prime}}}

\newcommand{\ametapr}{am_{\eta^{\prime}}}

\newcommand{\Nnoise}{N_{\rm noise}}

\newcommand{\Ninv}{N_{\rm inv}}

\newcommand{\alatt}{a}

\newcommand{\Orda}{{\mathrm O}(\alatt)}

\newcommand{\csw}{c_{\rm sw}}

\newcommand{\betakappa}{\left(\beta,\kappa\right)}

\newcommand{\ident}{\rlap{1}\kern0.2em\mathrm{l}}
\newcommand{\identn}[1]{\rlap{1}\kern0.2em\mathrm{l}_{#1}}
\newcommand{\identalbe}{\rlap{1}\kern0.2em\mathrm{l}_{\alpha\beta}}
\newcommand{\identcd}{\rlap{1}\kern0.2em\mathrm{l}_{cd}}
\newcommand{\identef}{\rlap{1}\kern0.2em\mathrm{l}_{cd}}

\newcommand{\Reals}{\rlap{1}\kern0.05em\mathrm{R}}

\newcommand{\pslash}{\rlap{$p$}\kern0.2em\mathrm{/}}
\newcommand{\qslash}{\rlap{$q$}\kern0.2em\mathrm{/}}
\newcommand{\Aslash}{\rlap{$A$}\kern0.2em\mathrm{/}}
\newcommand{\Aslashcd}{\rlap{$A$}\kern0.2em\mathrm{/}_{cd}}
\newcommand{\Dslash}{\rlap{$D$}\kern0.2em\mathrm{/}}
\newcommand{\Dslashcd}{\rlap{$D$}\kern0.2em\mathrm{/}_{cd}}
\newcommand{\Dslashcdalbe}
{\rlap{$D$}\kern0.2em\mathrm{/}_{cd\alpha\beta}}
\newcommand{\Dslashalbecd}
{\rlap{$D$}\kern0.2em\mathrm{/}_{\alpha\beta cd}}
\newcommand{\partslash}{\rlap{$\partial$}\kern0.2em\mathrm{/}}

\newcommand{\overview}{
\begin{table*}
\caption{Overview of full QCD simulations.  The lattice spacing $a$ is
fixed by the vector meson mass at the physical quark mass and
$\protect M_\rho=768.4$~MeV. The set of trajectories used for flavor
non-singlet hadrons \cite{cppacsspectrum,cppacsquarkmass} and
measurements of $G_{\rm disc}$ with a local source \cite{cppacspreliminary} is a
subset of those referred to under $N_{\rm Traj}$. }
\label{tab:overview}
\setlength{\tabcolsep}{1pc}
\begin{tabular}{ccccccccc}
\hline
$\beta$ & \lw{$L^3\times T$} & $a$ [fm] & \lw{$\kappa$} & 
\lw{$m_{\rm PS}/m_{\rm V}$} & \lw{$N_{\rm Traj}$} & 
\lw{$N_{\rm Skip}$} & \lw{$N_{\rm Smeared~Meas.}$} \\
$c_{SW}$ & & $La$ [fm] &&&&& \\
\hline
1.80 & \lw{$12^3{\times}24$} & 0.2150(22) &  
       0.1409 & 0.807(1) & 6530 & 10 & 651 \\
1.60 && 2.580(26)  & 0.1430 & 0.753(1) & 5240 & 10 & 521 \\
                 &&& 0.1445 & 0.694(2) & 7350 & 10 & 728 \\
                 &&& 0.1464 & 0.547(4) & 5250 & 10 & 407 \\ 
1.95 & \lw{$16^3{\times}32$} & 0.1555(17) & 
                     0.1375 & 0.804(1) & 7000 & 10 & 627 \\
1.53 && 2.489(27)  & 0.1390 & 0.752(1) & 7000 & 10 & 689 \\
                 &&& 0.1400 & 0.690(1) & 7000 & 10 & 689 \\
                 &&& 0.1410 & 0.582(3) & 5000 & 10 & 491 \\ 
2.10 & \lw{$24^3{\times}48$} & 0.1076(13) & 
                     0.1357 & 0.806(1) & 4000 & 5 & 799 \\
1.47 && 2.583(31)  & 0.1367 & 0.755(2) & 4000 & 5 & 776 \\
                 &&& 0.1374 & 0.691(3) & 4000 & 5 & 767 \\
                 &&& 0.1382 & 0.576(3) & 4000 & 5 & 785 \\ 
\hline
\end{tabular}
\end{table*}
}
\newcommand{\tabfitcomppswithfitrangesres}
{
\begin{table*}
\caption
{Fitted values of $\mpi$ and $\metapr$ for smeared source, with fit ranges.}
\label{tab:fitcomppswithfitrangesres}
\begin{center}
\begin{tabular}{cccccccc}
\hline
              $\beta$&$\kappa$  &   $a\mpi$  &  [$\tmin,\tmax$]  & 
 \multicolumn{4}{c}{$a\metapr$} \\
              & & &  & 
  Direct Fit  &[$\tmin,\tmax$]   &  Ratio Fit  &  [$\tmin,\tmax$] \\
\hline
\hline
  $\beta=1.8$&$ \kappa=0.1409$  &  1.155(1)  &     [5,12]  & 
             1.182(10)  &      [1,4]  &              1.218(8)  &      [1,4] \\
             &$ \kappa=0.1430$  &  0.984(1)  &     [6,12]  & 
             1.019(16)  &      [1,4]  &             1.057(11)  &      [1,5] \\
             &$ \kappa=0.1445$  &  0.821(1)  &     [6,12]  & 
             0.921(15)  &      [1,5]  &             0.937(15)  &      [1,5] \\
             &$ \kappa=0.1464$  &  0.532(2)  &     [6,12]  & 
             0.755(36)  &      [1,4]  &             0.769(35)  &      [1,5] \\
 $\beta=1.95$&$ \kappa=0.1375$  &  0.895(1)  &     [7,16]  & 
             0.960(13)  &      [1,4]  &             0.957(12)  &      [1,4] \\
             &$ \kappa=0.1390$  &  0.729(1)  &     [7,16]  & 
             0.846(13)  &      [1,5]  &             0.823(13)  &      [1,5] \\
             &$ \kappa=0.1400$  &  0.595(1)  &     [6,16]  & 
             0.754(12)  &      [1,4]  &             0.716(11)  &      [1,5] \\
             &$ \kappa=0.1410$  &  0.427(1)  &     [6,16]  & 
             0.705(24)  &      [1,4]  &             0.653(22)  &      [1,5] \\
  $\beta=2.1$&$ \kappa=0.1357$  &  0.630(1)  &    [10,24]  & 
              0.654(9)  &      [1,4]  &             0.680(10)  &      [1,4] \\
             &$ \kappa=0.1367$  &  0.516(1)  &    [10,24]  & 
              0.598(9)  &      [1,4]  &              0.602(9)  &      [1,5] \\
             &$ \kappa=0.1374$  &  0.424(1)  &    [10,24]  & 
             0.528(12)  &      [1,5]  &             0.515(11)  &      [1,5] \\
             &$ \kappa=0.1382$  &  0.295(1)  &    [10,24]  & 
             0.450(18)  &      [1,5]  &             0.426(18)  &      [1,5] \\
\hline
\end{tabular}
\end{center}
\end{table*}
}
\newcommand{\tabchiralcomppsres}
{
\begin{table}[h]
\caption
{Comparison of $\etapr$ mass at the physical point 
from Eqs. (\ref{eq:NGB}) (NGB fit) and 
(\ref{eq:nonNGB}) (non-NGB fit).}
\label{tab:chiralcomppsres}
\begin{center}
\begin{tabular}{ccc}
\hline
$\beta$     &\multicolumn{2}{c}{$\metapr({\rm GeV})$}                 \\
            &  NGB                &  non-NGB              \\
\hline
\hline
        1.8  &            0.509(39)    &                   0.589(24) \\
       1.95  &            0.714(36)    &                   0.766(26) \\
        2.1  &            0.709(41)    &                   0.764(30) \\
\hline
\end{tabular}
\end{center}
\end{table}
}
\newcommand{\tabsysone}
{
\begin{table}[h]
\caption
{Systematic variations in \ix{\metapr} over fit forms for chiral and continuum extrapolations.}
\label{tab:sysone}
\begin{center}
\begin{tabular}{cc}
\hline
                                     &  $\metapr$(GeV) \\
\hline
\hline
                     central value   &       0.960(87) \\
\hline
 quadratic continuum extrapolation  &       0.819(50) \\
  constant continuum extrapolation  &       0.712(27) \\
\hline
        non-N. G. B. chiral extrapolation  &       0.997(61) \\
\hline
\end{tabular}
\end{center}
\end{table}
}
\newcommand{\width}{8cm}
\newcommand{\figpismearcomp}
{
\begin{figure}
\begin{center}
\includegraphics[width=\width,clip]{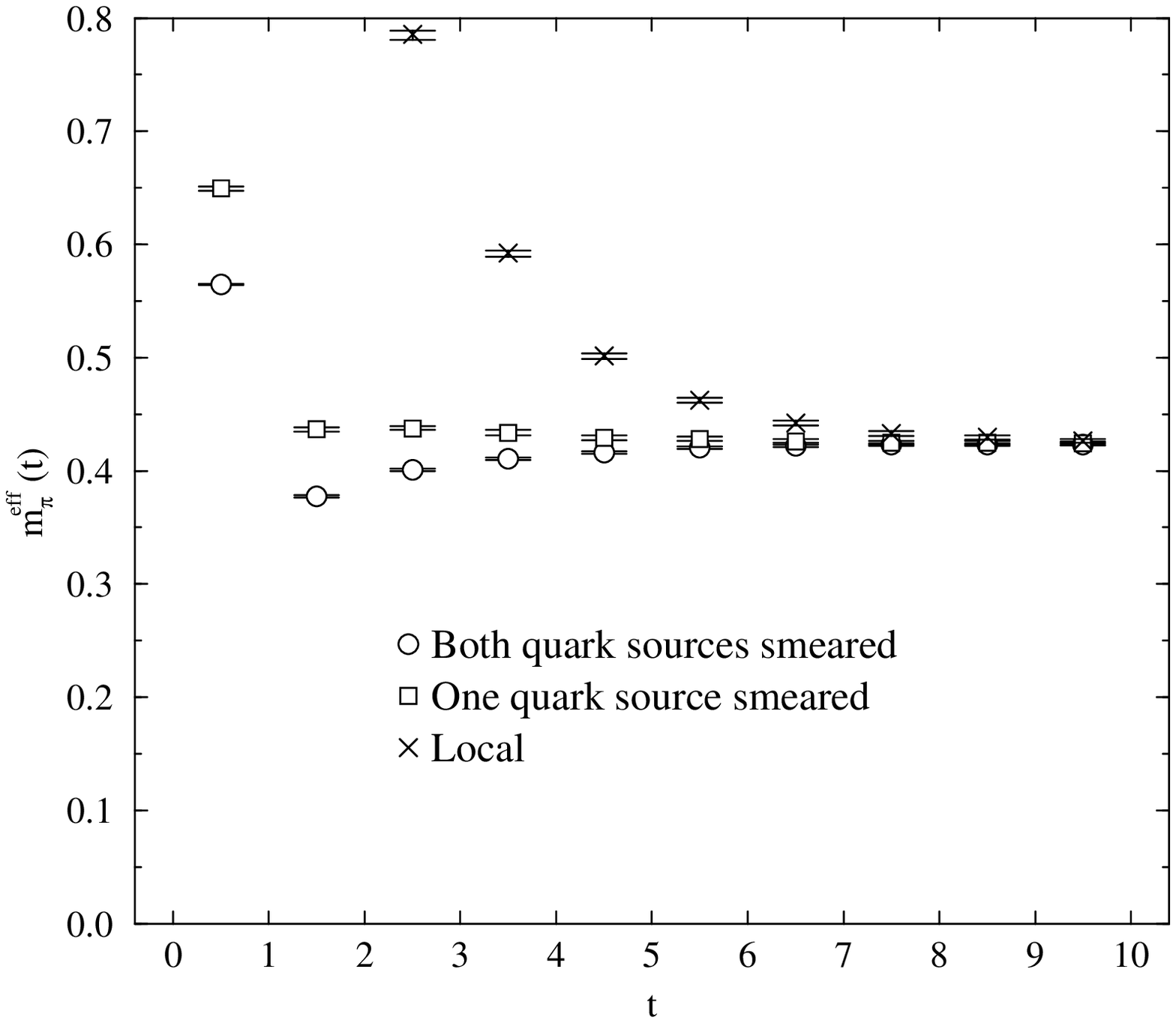}
\caption
{Comparison between smearing schemes for \ix{\pi} effective mass for \ix{\beta=2.1} and \ix{\kappa=0.1374} on a \ix{24^3\times48} lattice.}
\label{fig:pismearcomp}
\end{center}
\end{figure}
}
\newcommand{\figetasmearcomp}
{
\begin{figure}
\begin{center}
\includegraphics[width=\width,clip]{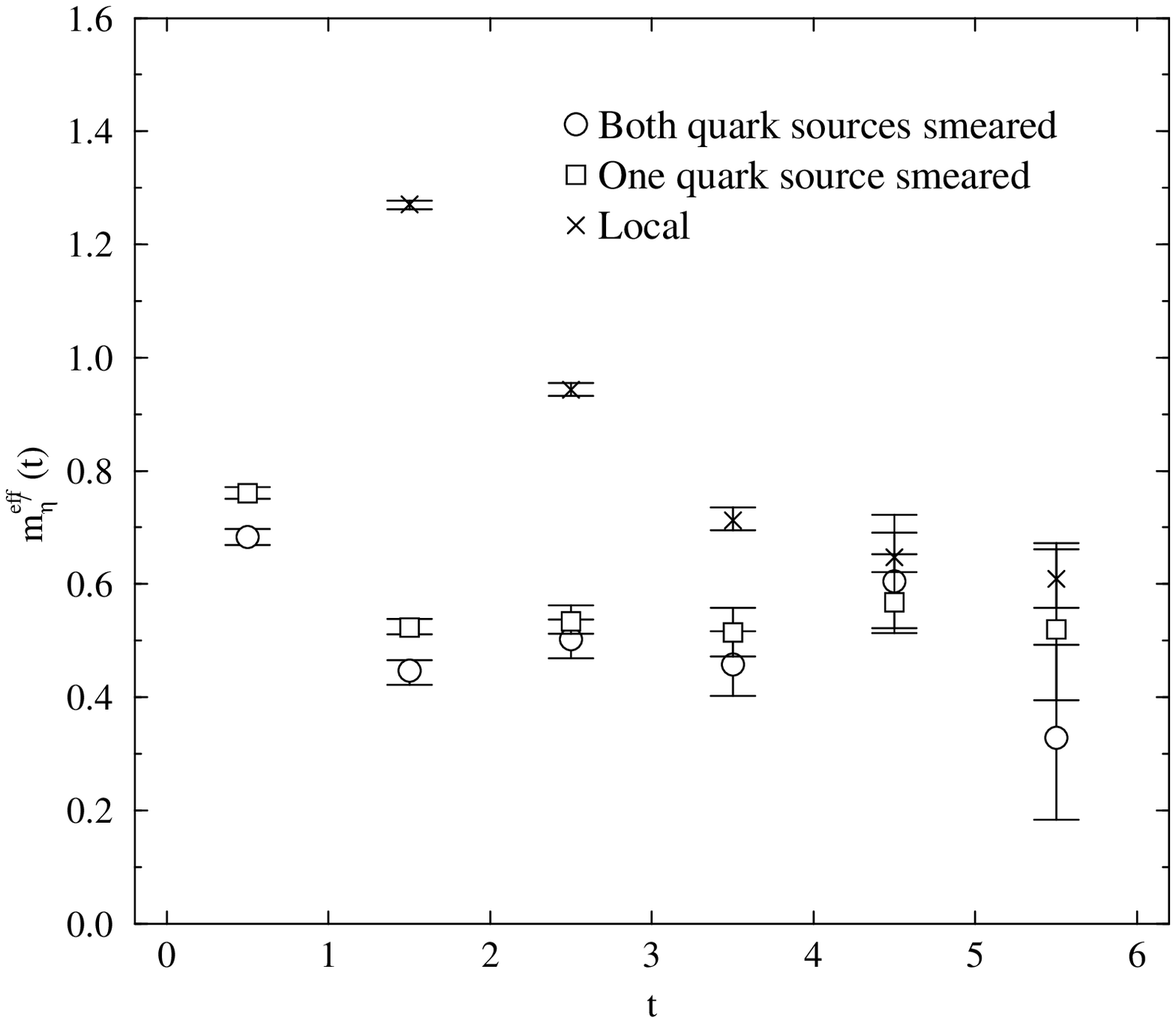}
\caption
{Comparison between smearing schemes for \ix{\etapr} effective mass for \ix{\beta=2.1} and \ix{\kappa=0.1374} on a \ix{24^3\times48} lattice.}
\label{fig:etasmearcomp}
\end{center}
\end{figure}
}
\newcommand{\fignoisedependence}
{
\begin{figure}
\begin{center}
\includegraphics[width=\width,clip]{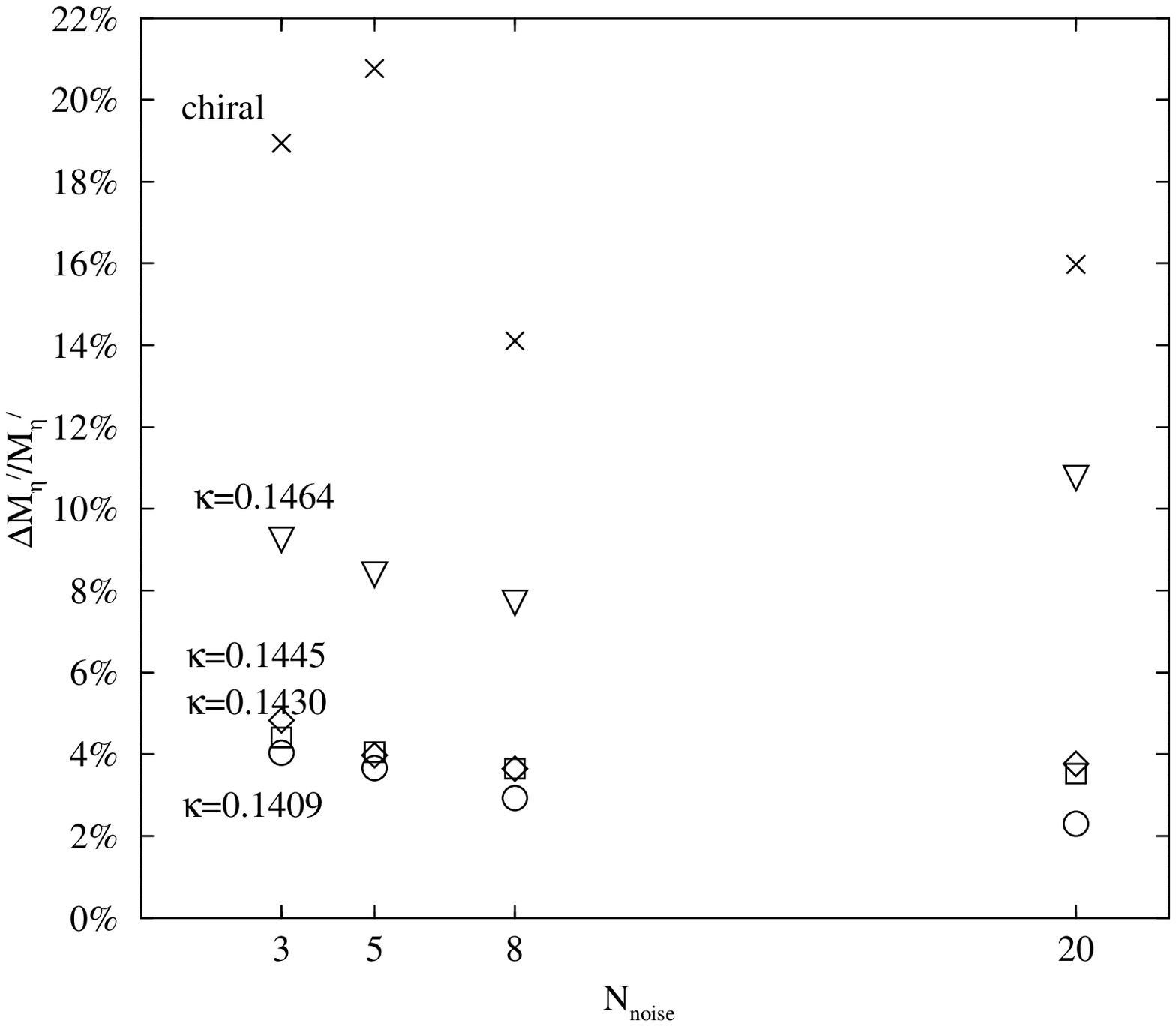}
\caption
{The same as Fig.~\protect\ref{fig:noisedependenceII} for relative error
of \ix{\metapr}. Data at the chiral limit are also shown by crosses.}
\label{fig:noisedependence}
\end{center}
\end{figure}
}
\newcommand{\fignoisedependenceII}
{
\begin{figure}
\begin{center}
\includegraphics[width=\width,clip]{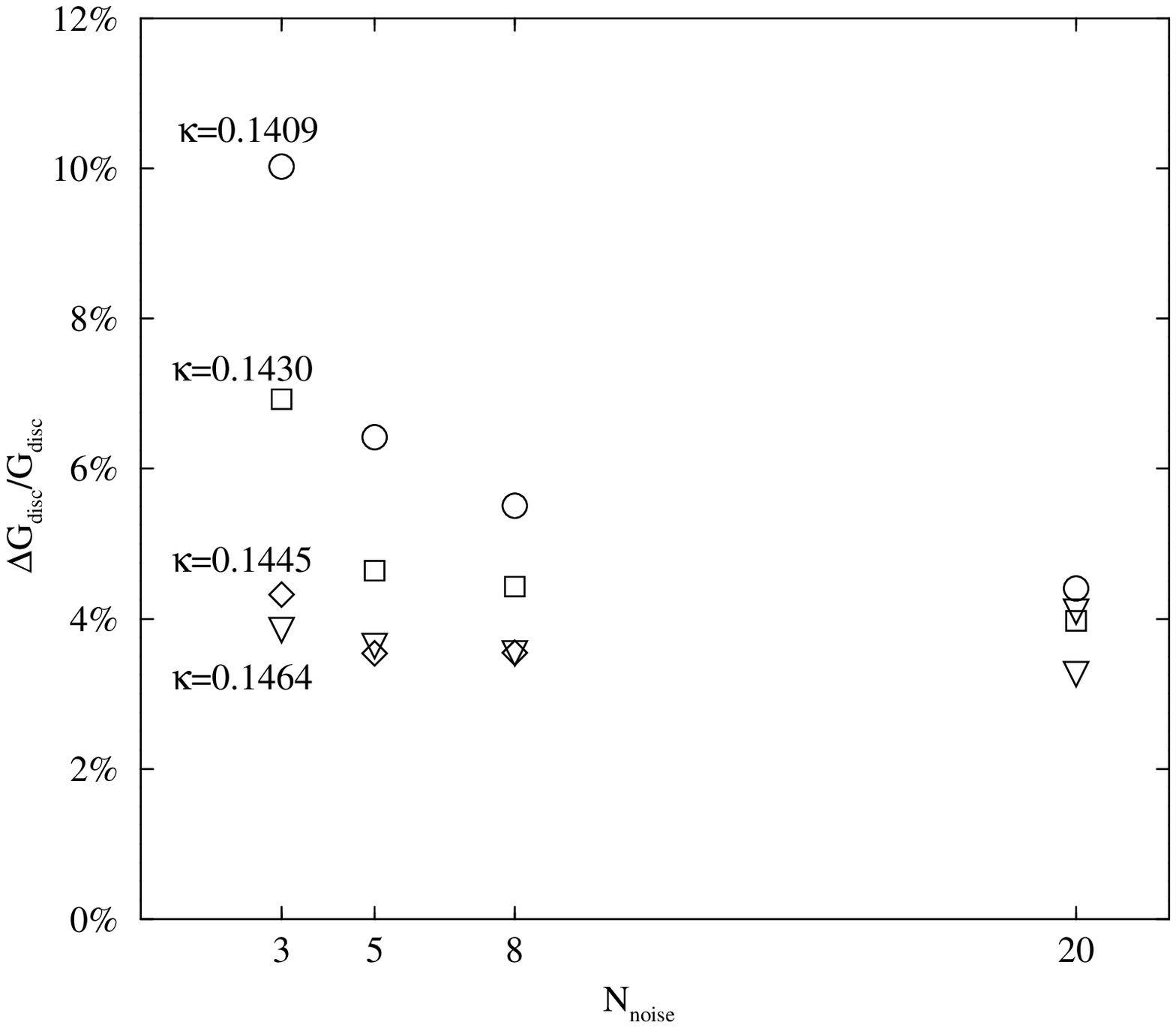}
\caption{Relative error of one-quark smeared \ix{\Gdisc(t=2)} versus
\ix{\Nnoise} at \ix{\beta=1.8}. }
\label{fig:noisedependenceII}
\end{center}
\end{figure}
}
\newcommand{\figetaefflotwtw}
{
\begin{figure}
\begin{center}
\includegraphics[width=\width,clip]{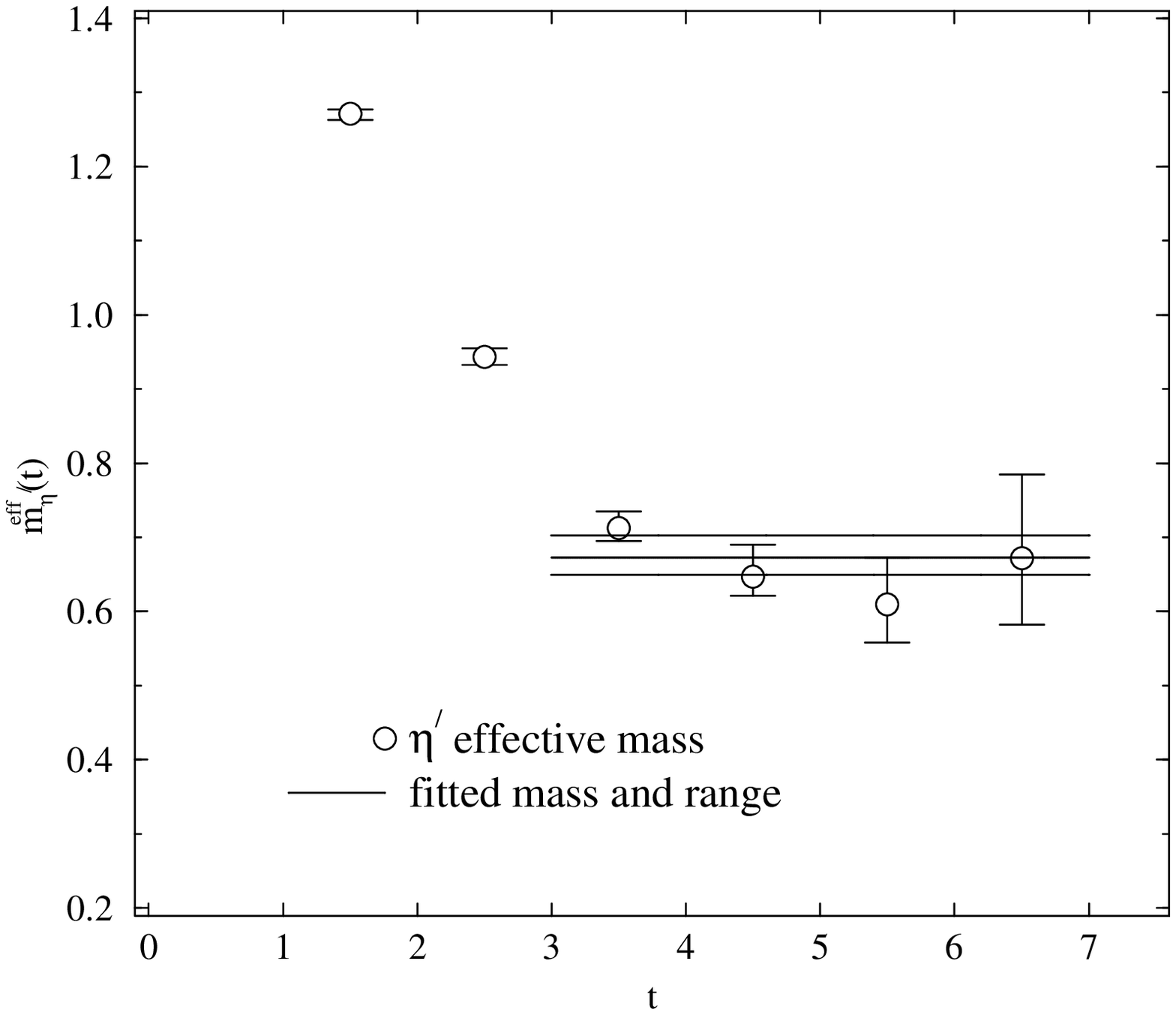}
\caption
{\ix{\etapr} effective mass \ix{\metapr^{\rm eff}(t)}
obtained from local
source at \ix{\beta=2.1} and \ix{\kappa=0.1374} on 
a \ix{24^3\times 48} lattice.
} 
\label{fig:etaefflotwtw}
\end{center}
\end{figure}
}
\newcommand{\figratlotwtw}
{
\begin{figure}
\begin{center}
\includegraphics[width=\width,clip]{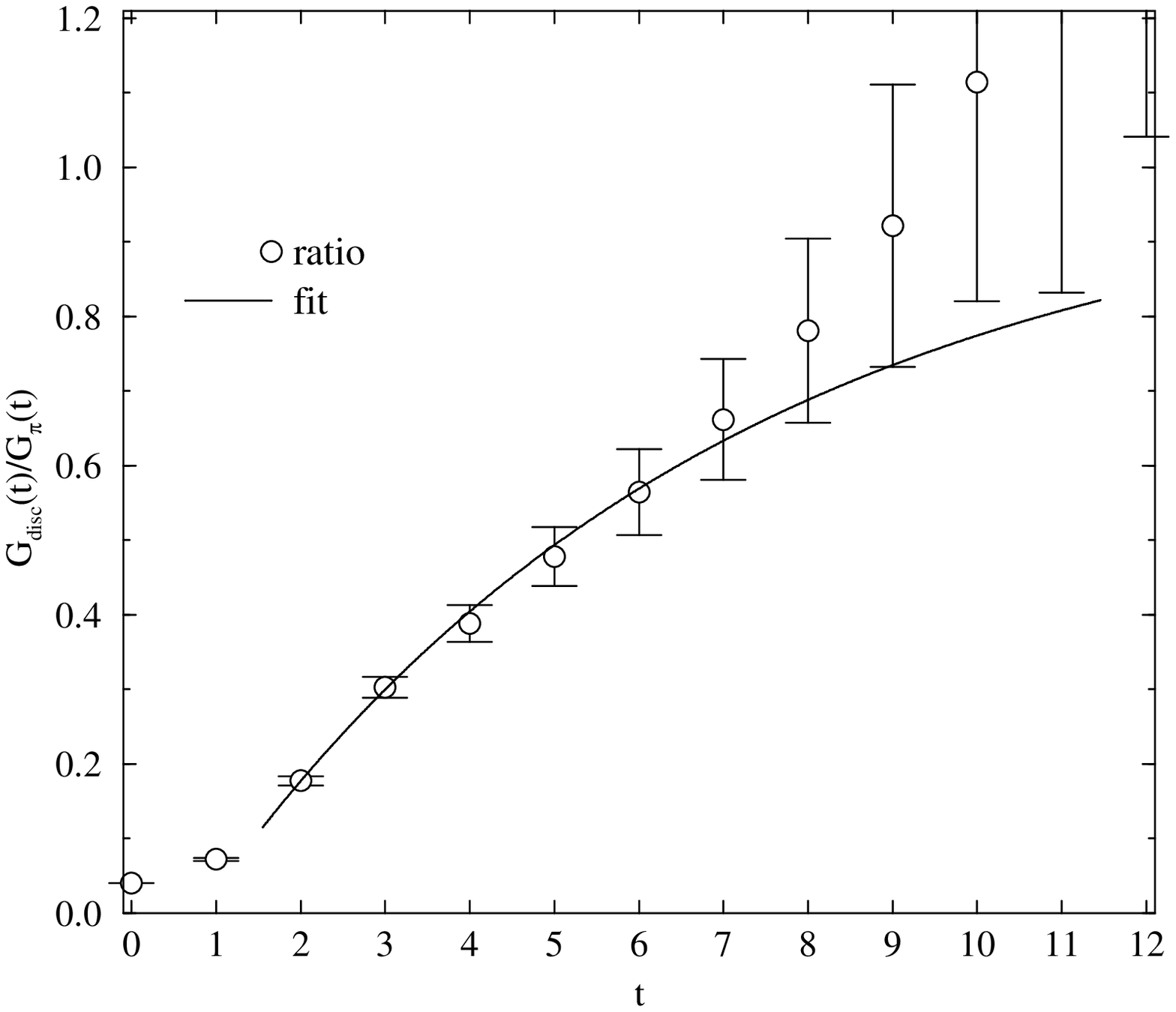}
\caption
{Ratio \ix{\Gdisc(t)/\Gpi(t)} from local source for \ix{\beta=2.1} and
\ix{\kappa=0.1374} on a \ix{24^3\times 48} lattice.} 
\label{fig:ratlotwtw}
\end{center}
\end{figure}
}
\newcommand{\figetaeffsmtwtw}
{
\begin{figure}
\begin{center}
\includegraphics[width=\width,clip]{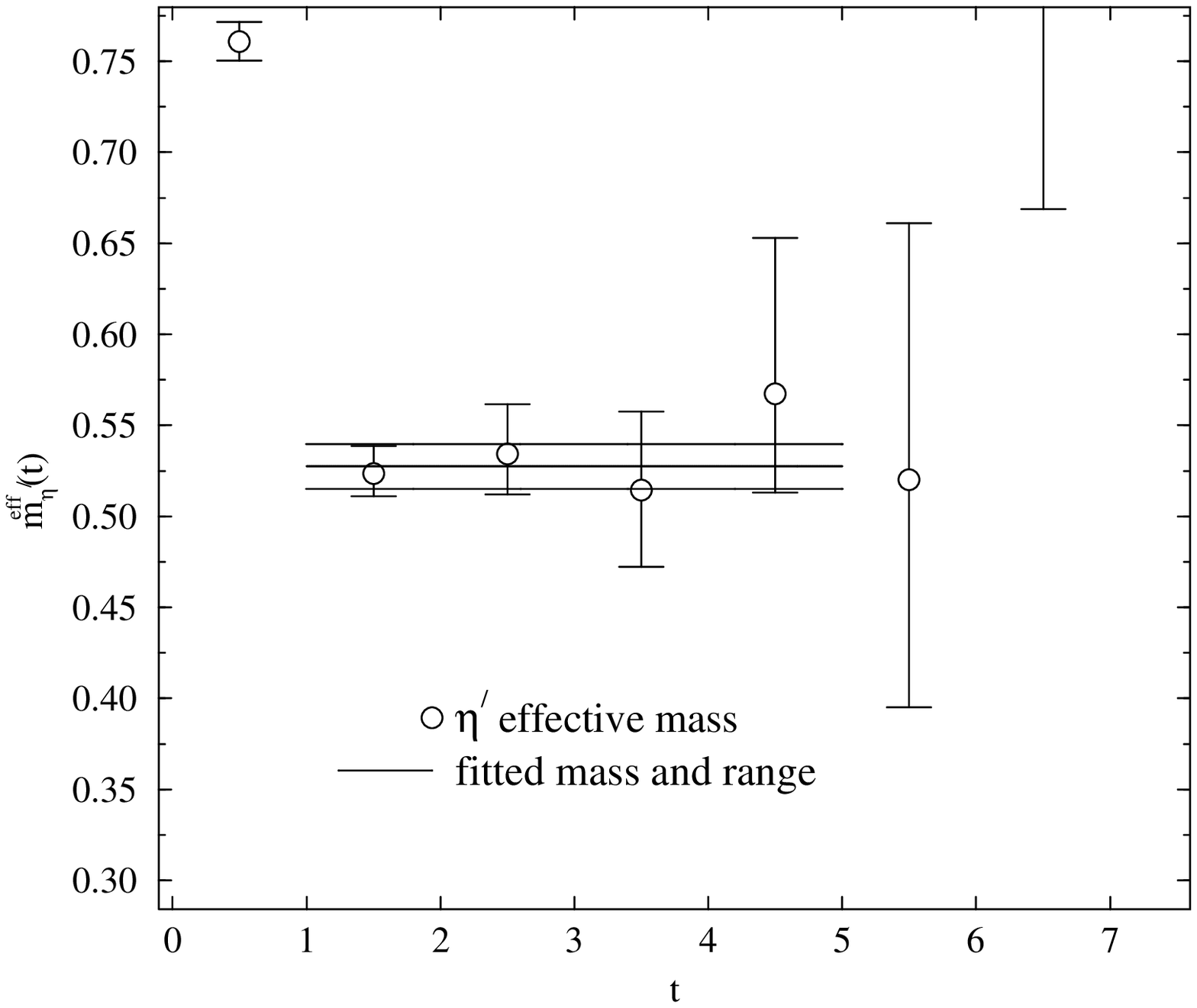}
\caption
{\ix{\etapr} effective mass \ix{\metapr^{\rm eff}(t)} from smeared
source for \ix{\beta=2.1} and \ix{\kappa=0.1374} on a 
\ix{24^3\times 48} lattice.} 
\label{fig:etaeffsmtwtw}
\end{center}
\end{figure}
}
\newcommand{\figratsmtwtw}
{
\begin{figure}
\begin{center}
\includegraphics[width=\width,clip]{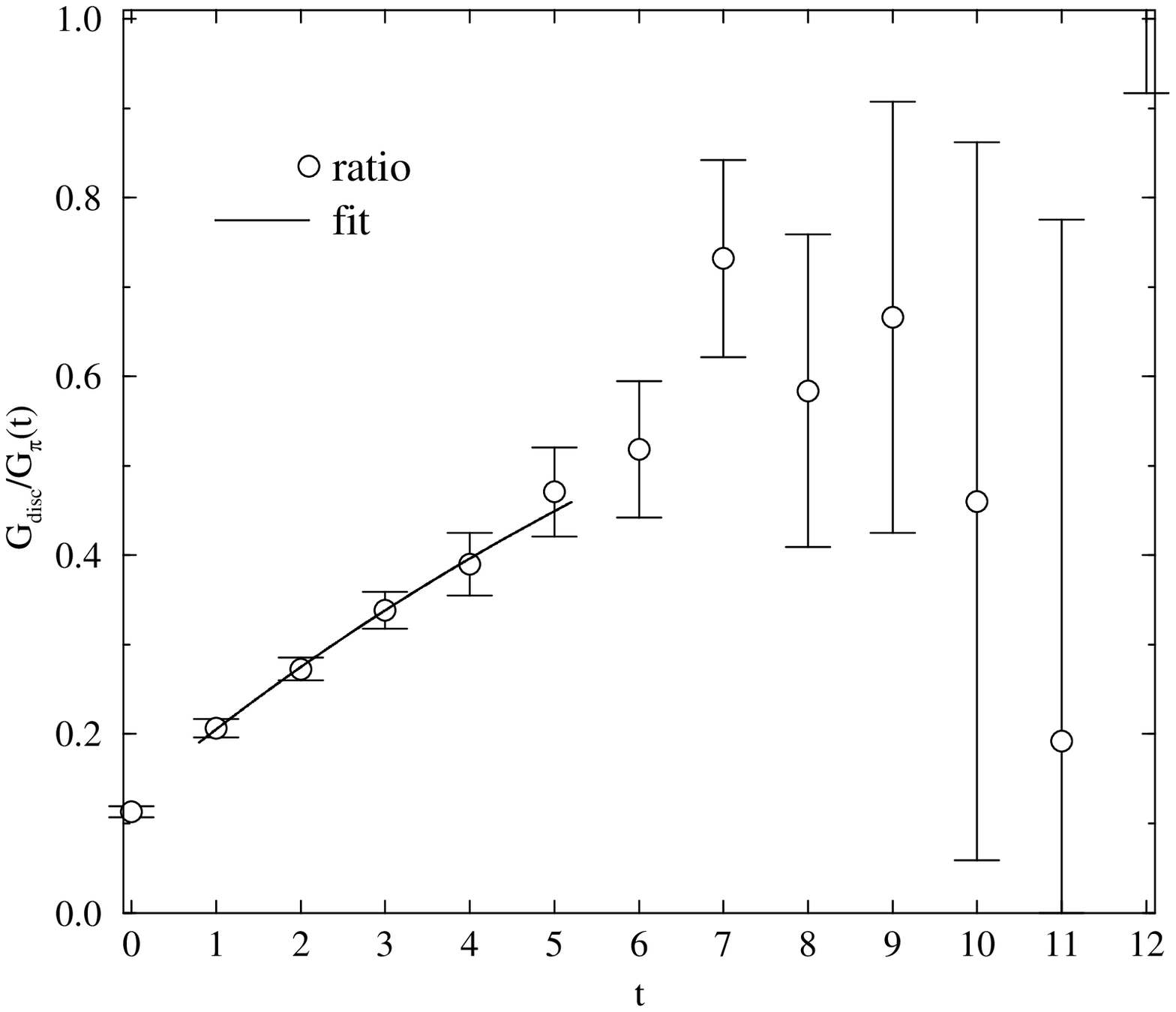}
\caption
{Ratio \ix{\Gdisc(t)/\Gpi(t)} from smeared source for \ix{\beta=2.1}
and \ix{\kappa=0.1374} on a \ix{24^3\times 48} lattice.} 
\label{fig:ratsmtwtw}
\end{center}
\end{figure}
}
\newcommand{\figetaeffsmze}
{
\begin{figure}
\begin{center}
\includegraphics[width=\width,clip]{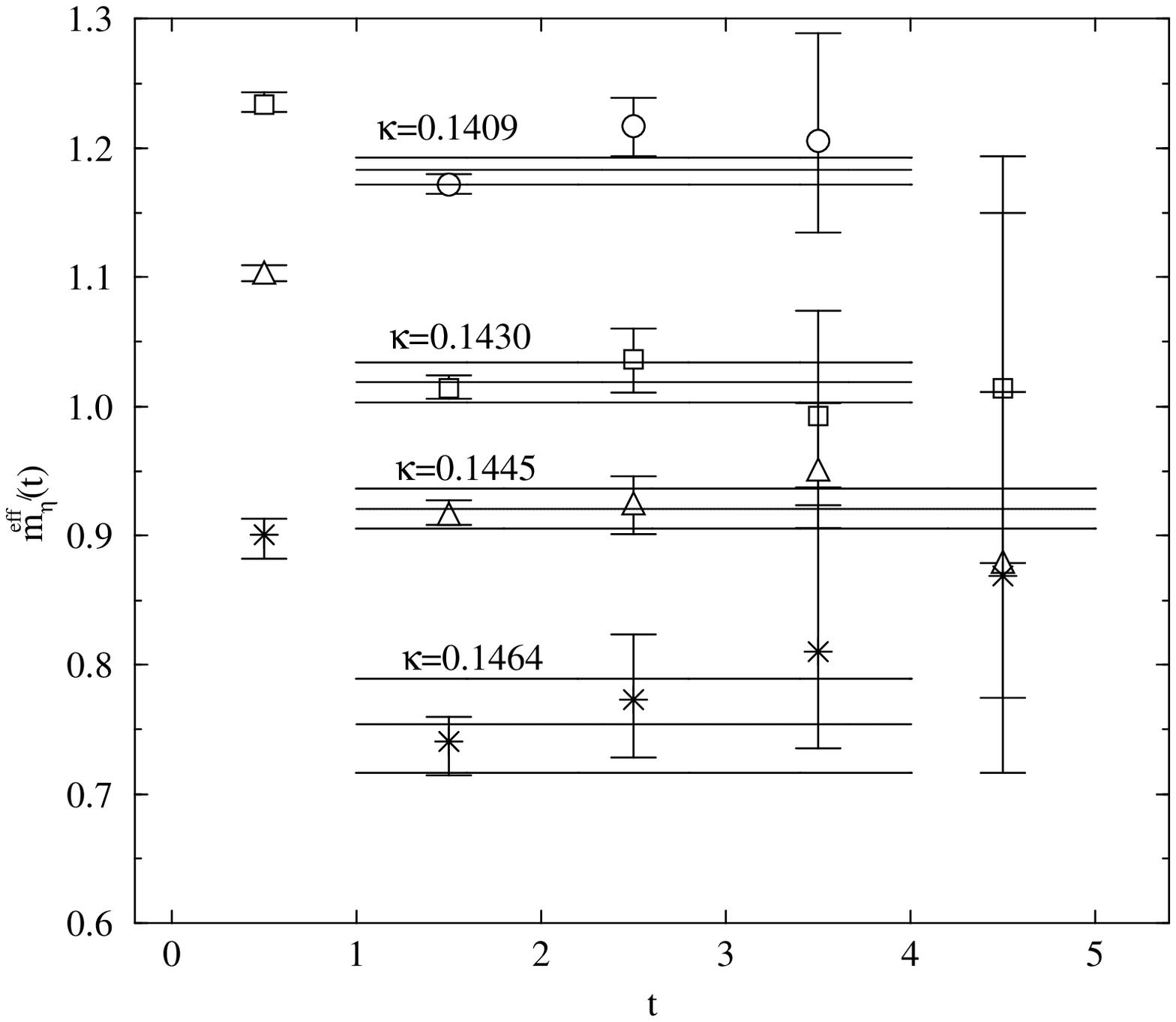}
\caption
{\ix{\etapr} effective masses \ix{\metapr^{\rm eff}(t)} from one-quark
smeared sources for \ix{\beta}=1.8 on a \ix{12^3\times 24} lattice.} 
\label{fig:etaeffsmze}
\end{center}
\end{figure}
}
\newcommand{\figetaeffsmon}
{
\begin{figure}
\begin{center}
\includegraphics[width=\width,clip]{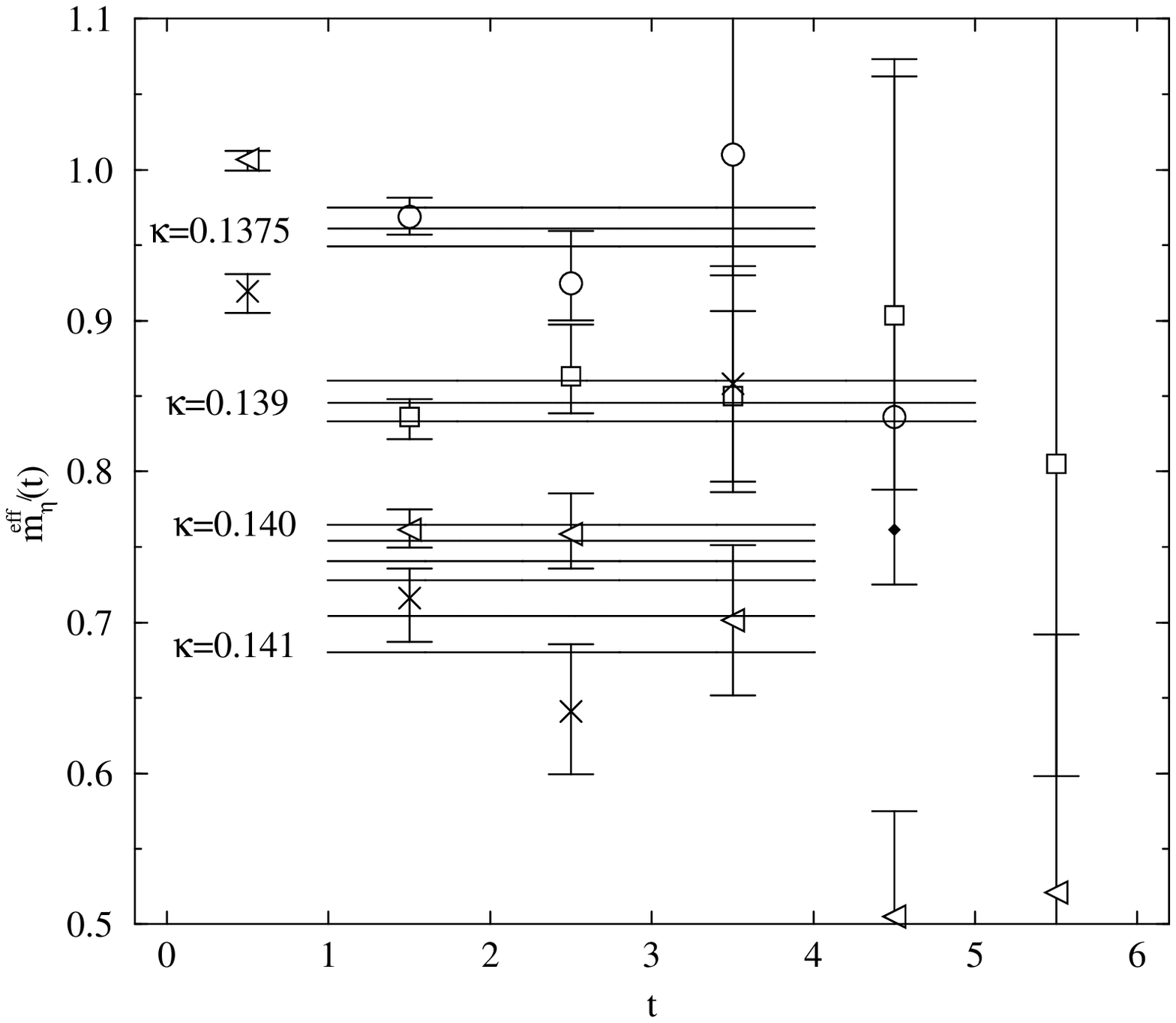}
\caption
{\ix{\etapr} effective masses \ix{\metapr^{\rm eff}(t)} from one-quark
smeared sources for \ix{\beta}=1.95 on a \ix{16^3\times 32} lattice.}
\label{fig:etaeffsmon}
\end{center}
\end{figure}
}
\newcommand{\figetaeffsmtw}
{
\begin{figure}
\begin{center}
\includegraphics[width=\width,clip]{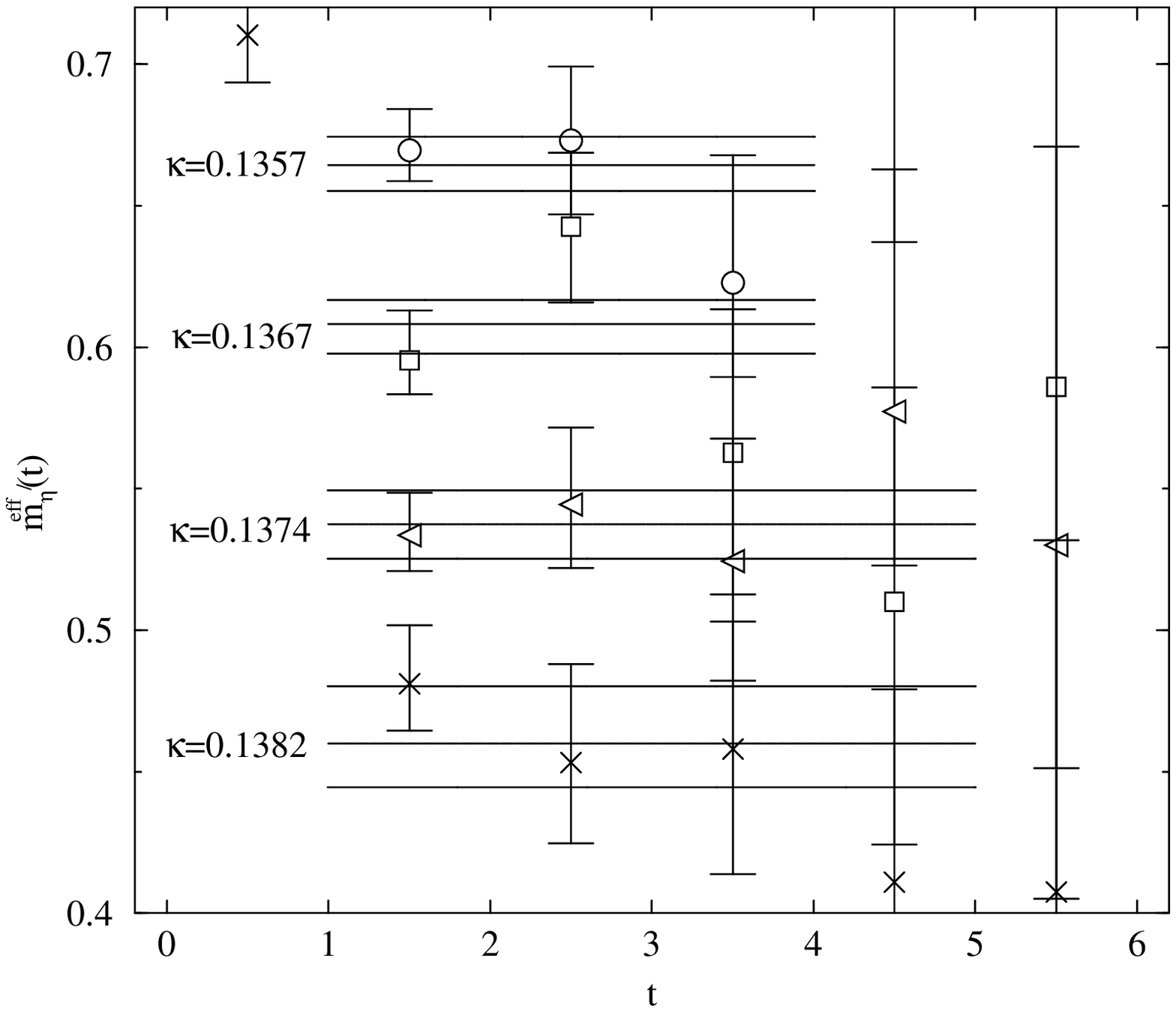}
\caption
{\ix{\etapr} effective masses \ix{\metapr^{\rm eff}(t)} from one-quark
smeared sources for \ix{\beta}=2.1 on a \ix{24^3\times 48} lattice.}
\label{fig:etaeffsmtw}
\end{center}
\end{figure}
}
\newcommand{\figchiralxze}
{
\begin{figure}
\begin{center}
\includegraphics[width=\width,clip]{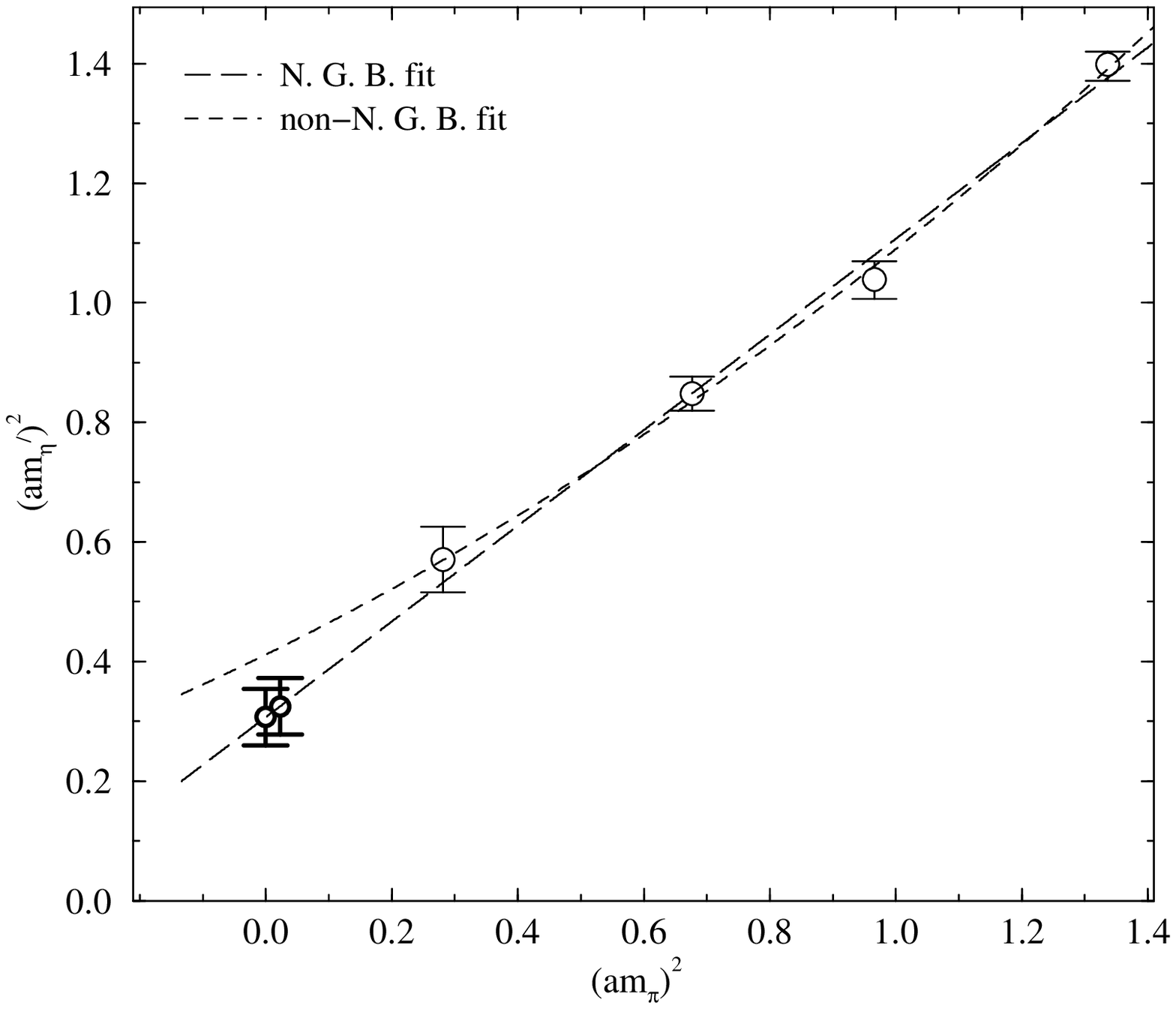}
\caption
{Chiral extrapolation for \ix{\beta=1.8} and smeared sources.
The two extrapolated points correspond to 
physical and to zero pion mass.}
\label{fig:chiralxze}
\end{center}
\end{figure}
}
\newcommand{\figchiralxon}
{
\begin{figure}
\begin{center}
\includegraphics[width=\width,clip]{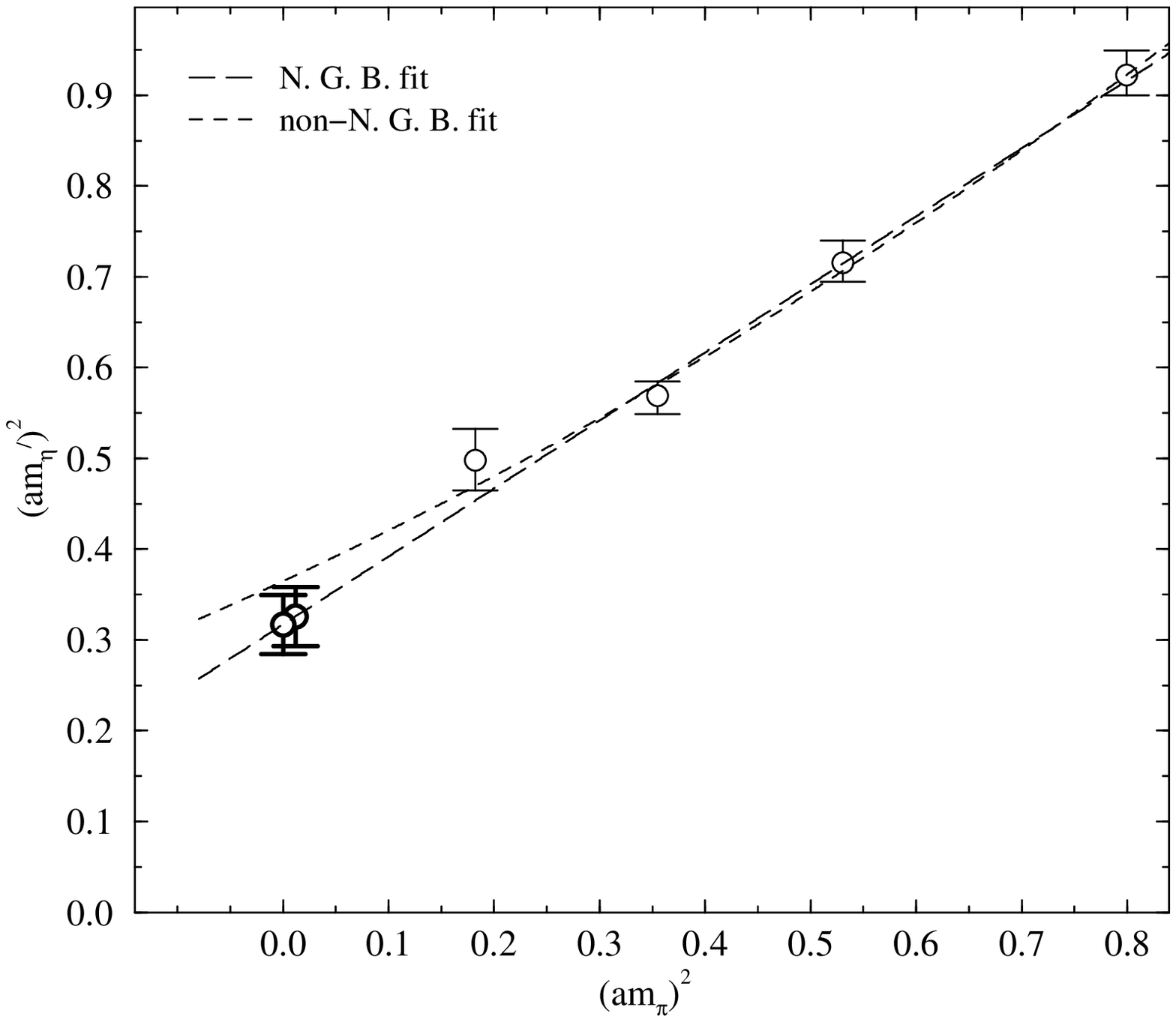}
\caption
{Chiral extrapolation for \ix{\beta=1.95} and smeared sources. The meaning of symbols is the same as in Fig.~\ref{fig:chiralxze}.}
\label{fig:chiralxon}
\end{center}
\end{figure}
}
\newcommand{\figchiralxtw}
{
\begin{figure}
\begin{center}
\includegraphics[width=\width,clip]{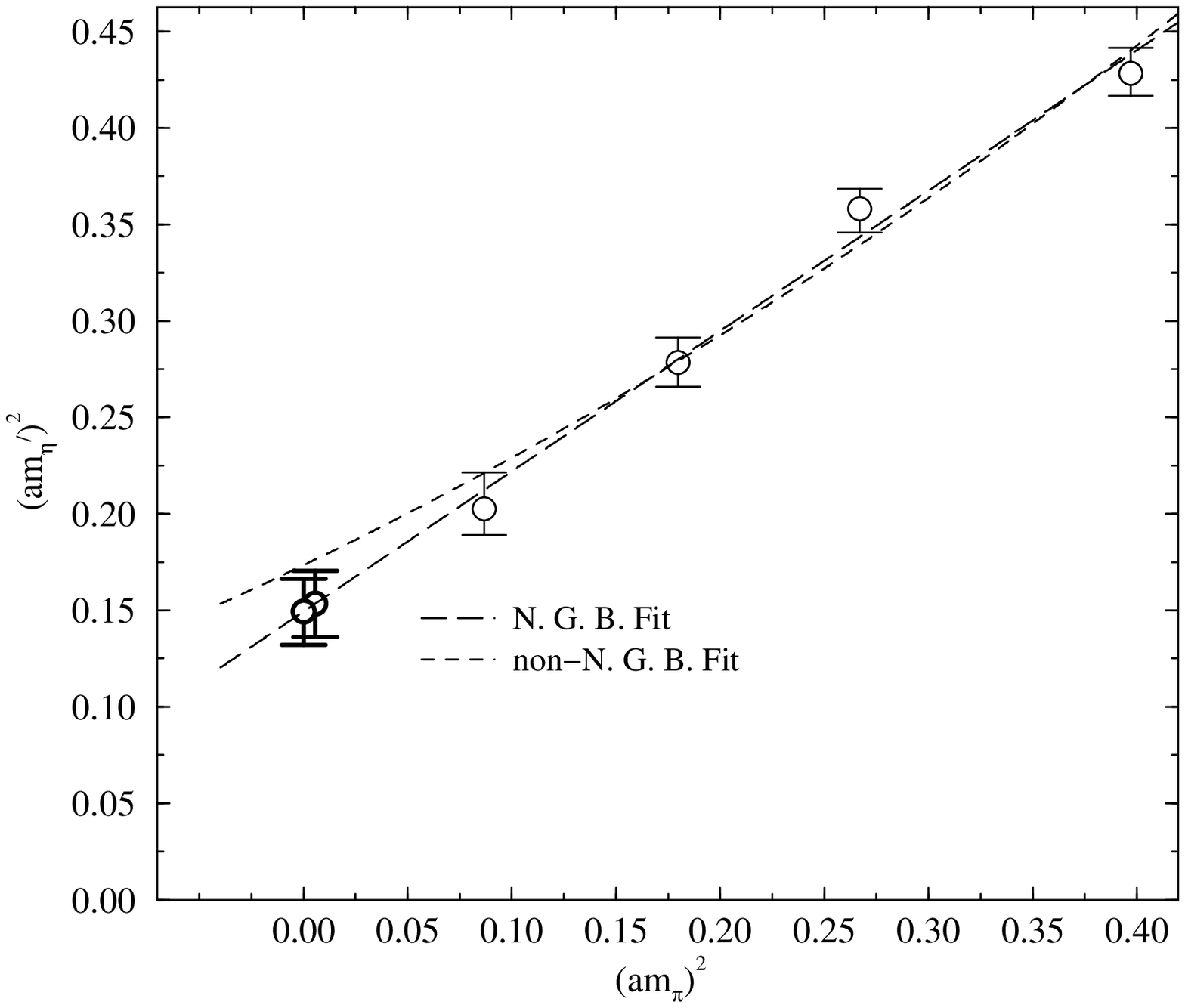}
\caption
{Chiral extrapolation for \ix{\beta=2.1}, smeared sources. The meaning of symbols is the same as in Fig.~\ref{fig:chiralxze}.}
\label{fig:chiralxtw}
\end{center}
\end{figure}
}
\newcommand{\figcontx}
{
\begin{figure}
\begin{center}
\includegraphics[width=\width,clip]{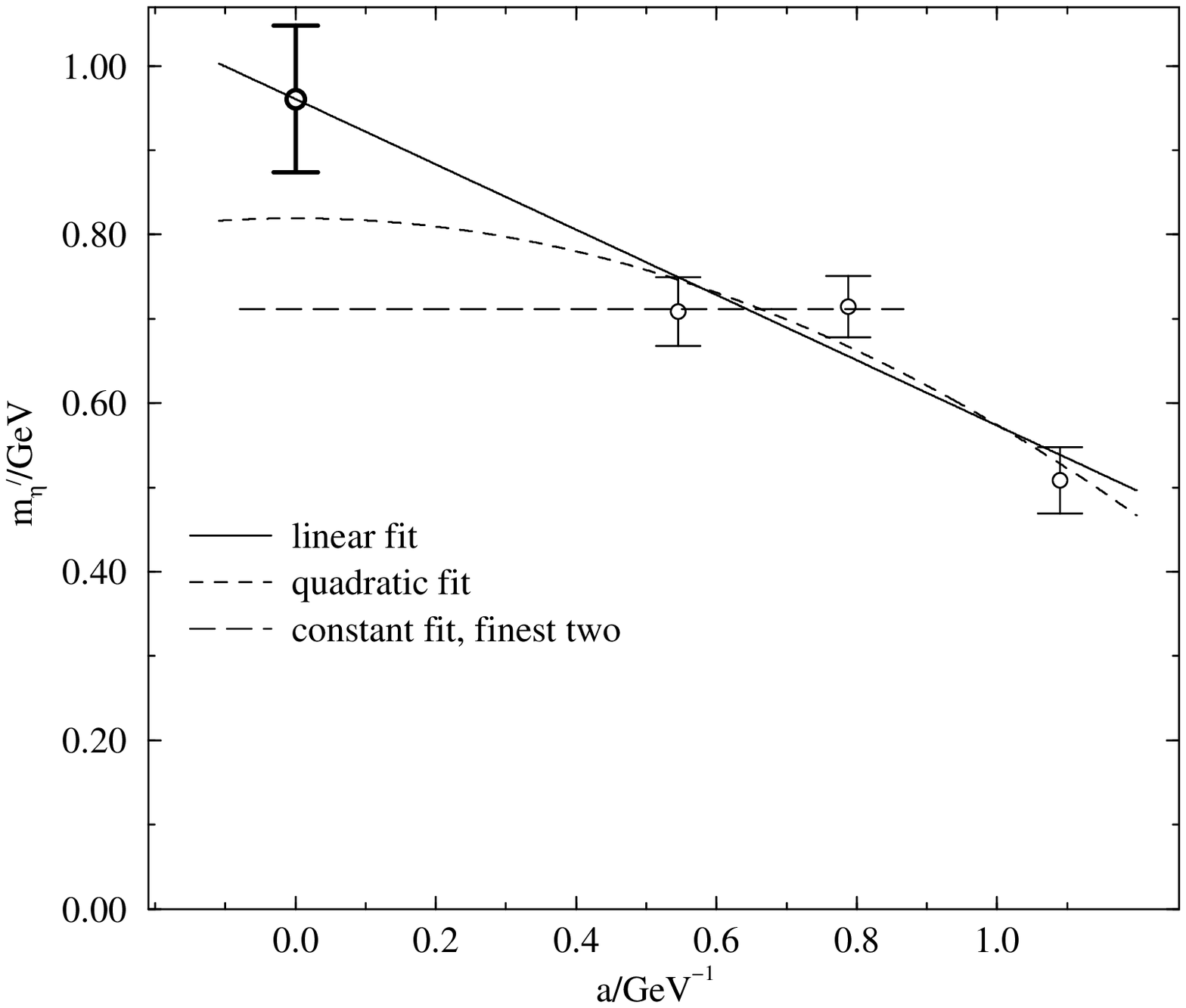}
\caption
{Continuum extrapolation of \ix{\etapr} meson mass obtained with smeared sources.}
\label{fig:contx}
\end{center}
\end{figure}
}
\newcommand{\figetacomp}
{
\begin{figure}
\begin{center}
\includegraphics[width=\width,clip]{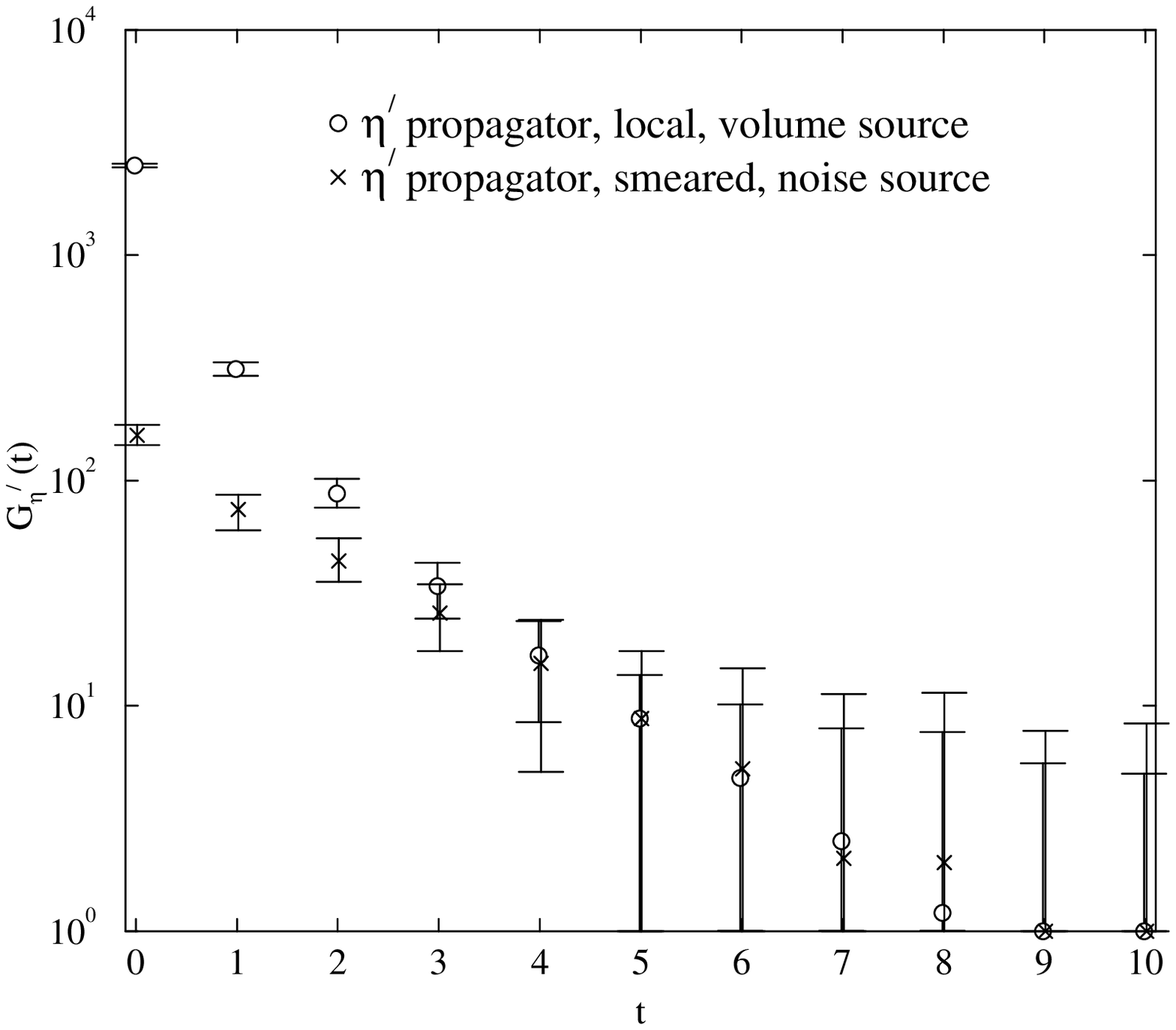}
\caption
{Comparison of \ix{\etapr} propagator for local and smeared source for \ix{\beta=2.1} and \ix{\kappa=0.1374} on a \ix{24^3\times 48} lattice.
The local source propagator has been normalized to be coincident with the smeared source propagator at t=5.}
\label{fig:etacomp}
\end{center}
\end{figure}
}
\newcommand{\figpieffectivemasscomp}
{
\begin{figure}
\begin{center}
\includegraphics[width=\width,clip]{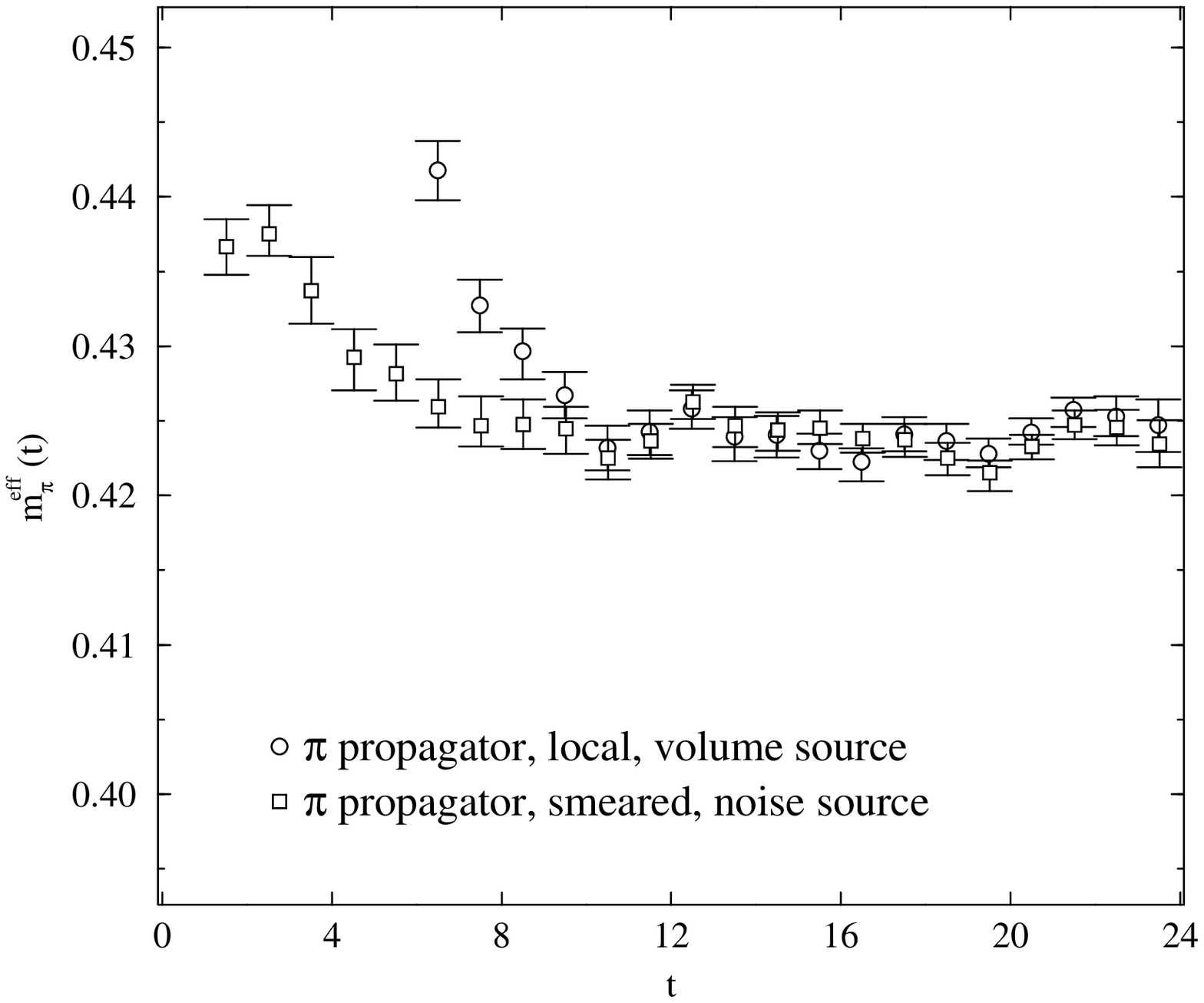}
\caption
{Comparison of pion effective mass for local and smeared source for \ix{\beta=2.1} and \ix{\kappa=0.1374} on a \ix{24^3\times 48} lattice.}
\label{fig:pieffectivemasscomp}
\end{center}
\end{figure}
}
\newcommand{\figetaeffectivemasscomp}
{
\begin{figure}
\begin{center}
\includegraphics[width=\width,clip]{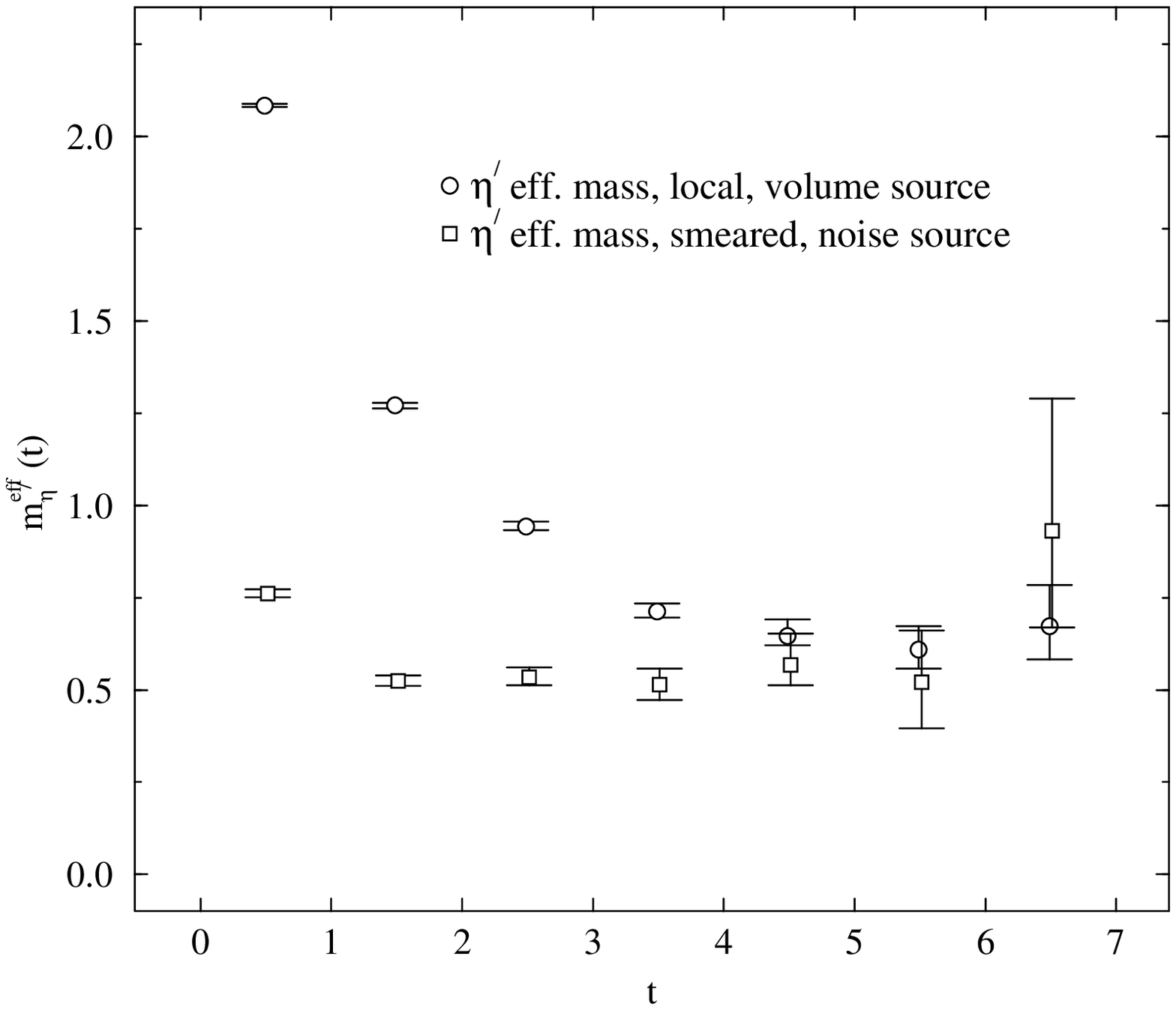}
\caption
{Comparison of \ix{\etapr} effective mass for local and smeared source for \ix{\beta=2.1} and \ix{\kappa=0.1374} on a \ix{24^3\times 48} lattice.}
\label{fig:etaeffectivemasscomp}
\end{center}
\end{figure}
}
\newcommand{\figdirratcomp}
{
\begin{figure}
\begin{center}
\includegraphics[width=\width,clip]{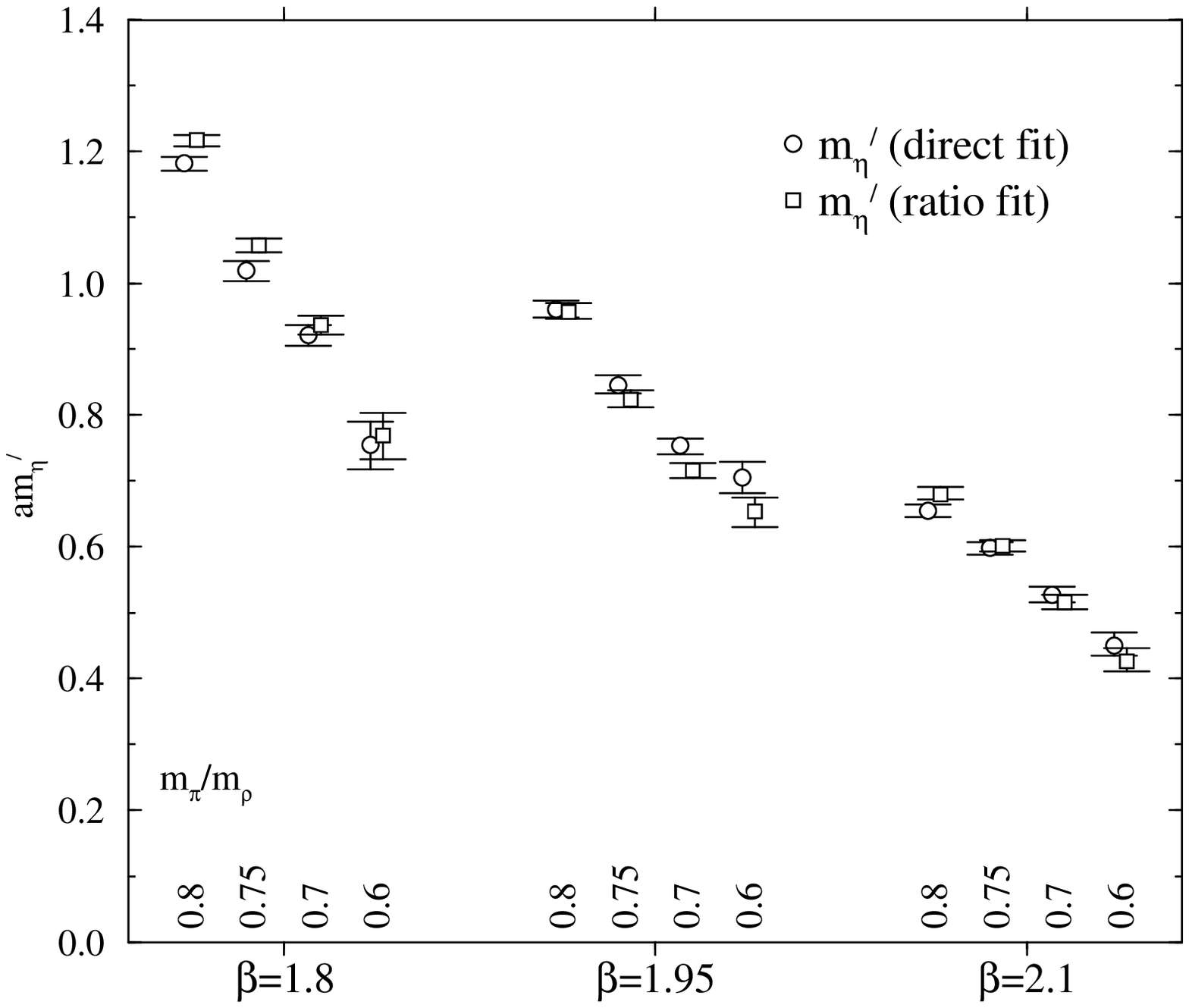}
\caption
{Comparison of \ix{\etapr} effective mass from direct and ratio fit (smeared source).}
\label{fig:dirratcomp}
\end{center}
\end{figure}
}
\begin{document}

\preprint{CCP-P-133}
\preprint{hep-lat/0211040}

\title{Flavor Singlet Meson Mass in the Continuum Limit 
in Two-Flavor Lattice QCD}
\author{V. I. Lesk$^1$, S. Aoki$^2$, 
R. Burkhalter$^{1,2}$
\footnote{present address: KPMG Consulting AG, Badenerstrasse 172, 8804 Z\"urich, Switzerland},
M. Fukugita$^3$, K.-I. Ishikawa$^{2,3}$, N. Ishizuka$^{1,2}$, Y. Iwasaki$^{1,2}$, K. Kanaya$^{1,2}$, Y. Kuramashi$^4$, M. Okawa$^4$, Y. Taniguchi$^2$, A. Ukawa$^{1,2}$, T. Umeda$^1$, T. Yoshie$^{1,2}$\\
(CP-PACS Collaboration)}
\affiliation{%
$^1$Center for Computational Physics, University of Tsukuba, Tsukuba, Ibaraki 305-8577, Japan\\
$^2$Institute of Physics, University of Tsukuba, Tsukuba, Ibaraki 305-8571, Japan\\
$^3$Institute for Cosmic Ray Research, University of Tokyo, Kashiwa 277-8582, Japan\\
$^4$High Energy Accelerator Research Organization (KEK),  Tsukuba, Ibaraki 305-0801, Japan
}%

\date{\today}

\begin{abstract}
We present results for the mass of the $\eta^\prime$ meson in the 
continuum limit for two-flavor lattice QCD, 
calculated on the CP-PACS computer, 
using a renormalization group improved gauge action, and Sheikoleslami
and Wohlert's fermion action with tadpole-improved $\csw$.
Correlation functions are measured at three values of the coupling 
$\beta$ corresponding to the lattice spacing $a\approx 0.22, 0.16, 0.11$~fm 
and for four values of the quark mass parameter $\kappa$ corresponding to 
$m_\pi/m_\rho\approx 0.8, 0.75, 0.7, 0.6$. 
For each $(\beta, \kappa)$ pair, 400-800 gauge configurations are used. 
The two-loop diagrams are evaluated using a noisy source method.
We calculate $\eta^\prime$ propagators using both smeared and local sources,
and find that excited state contaminations are much reduced by smearing. 
A full analysis for the smeared propagators gives 
$\metapr = 0.960(87)^{+0.036}_{-0.248}$ GeV in the continuum limit, 
where the second error represents the systematic uncertainty coming 
from varying the functional form for chiral and continuum extrapolations.

\end{abstract}

\pacs{11.15.Ha,12.38.Gc}

\maketitle

\section{Introduction}
\label{sec:introduction}

The large mass of the $\eta^\prime$ meson relative to members of the
pseudoscalar octet has been an outstanding problem in low-energy hadron
spectroscopy for some time~\cite{weinberg,metathooft,wittenveneziano}.  
A number of lattice QCD calculations has been carried out~\cite{fukugitaetal,
itohetal,kuramashietal,duncanetal,kilcup,mcneile,sesam}
to reproduce this feature, and to try to understand how it arises.
These simulations, however, were made
either with quenched configurations~\cite{fukugitaetal,itohetal,kuramashietal,
duncanetal} or, where full QCD was employed~\cite{kilcup,mcneile,sesam}, 
at a single lattice spacing.  
In this
article we report on a two-flavor full QCD calculation of the flavor
singlet meson mass including the continuum extrapolation. The calculation is 
made on a set of gauge configurations previously 
generated for a study of light hadron physics, the results of which have been reported in
Ref.~\cite{cppacsspectrum} for meson and baryon spectra and
in Ref.~\cite{cppacsquarkmass} for light quark masses.
Three values of lattice spacing in the range $a\approx 0.22-0.11$~fm, 
and four values of quark mass covering $m_\pi/m_\rho\approx 0.8-0.6$ are used.

The main computational challenge in this work lies in the estimation
of the the double quark loop diagram contribution to the $\eta^\prime$
propagator $G_{\eta^\prime}(t)$, 
for which the relative error \ix{\Delta \Getapr(t) /
\Getapr(t)} increases quickly with \ix{t}, 
the time separation from the source.
If the error becomes large at a time slice \ix{t_{\rm err}} 
less than the first time slice of the plateau in the effective mass 
\ix{\tmin}, it becomes undesirable to fit the \ix{\Getapr} directly.  
In such circumstances the ratio $G_{\eta^\prime}(t)/G_\pi(t)$ of 
$\eta^\prime$ to $\pi$ propagator has been used to try to cancel 
the effects of excited state contributions \cite{kilcup,cppacspreliminary}. 
In this work we set out to ensure a plateau, 
using an exponential-like smeared 
source and the method of \ix{U(1)} noise source \cite{noisenoise} to 
decrease \ix{\tmin}. The smearing technique has been previously employed in 
Refs.~\cite{duncanetal,kilcup,mcneile,sesam}.
We also calculate the disconnected contribution with a local source using the
method of volume source without gauge fixing \cite{kuramashietal}.

Preliminary results with the local source have been reported in
Ref.~\cite{cppacspreliminary}. Here we present full results, and carry
our analysis through in the case of the smeared source, 
where plateaux are achieved.

The organization of this article is as follows. Details of numerical
calculations are described in Sec.~\ref{sec:numerical}. 
In Sec.~\ref{sec:analysis}, we give a full account of our analysis
procedure and results including estimates of systematic errors arising
from  chiral and continuum extrapolations. Conclusions are given in
Sec.~\ref{sec:conclusion}.

\section{Numerical Calculations}\label{sec:numerical}

\subsection{Action and Configuration}
\label{subsec:actionandconfigurationgeneration}

\overview

We use full QCD configurations for two flavors of dynamical quarks. 
In order to reduce discretization error, these configurations are 
generated with a renormalization-group (RG) improved gauge action 
and a mean field improved clover quark action.  
The RG-improved gauge action~\cite{iwasakiaction} has the form 
\begin{equation}
S_{\rm RG} = 
\frac{\beta}{6} \left\{\;c_0\sum_{x,\mu <\nu} W_{\mu\nu}^{1\times 1}(x) 
+c_1\sum_{x,\mu ,\nu} W_{\mu\nu}^{1\times 2}(x)\right\},
\label{eq:RGAct} 
\end{equation}
where $W^{i\times j}$ are Wilson loops with size $i\times j$, 
$c_1 = -0.331$, and $c_0=1-8c_1$.
For the clover quark action~\cite{cloveraction}, we set  
the coefficient $c_{\rm SW}=P^{-3/4}$, where $P$ is the
plaquette value calculated in perturbation theory at one loop as
$P=1-0.8412\beta^{-1}$. 

A summary of the parameters and statistics is given in
Table~\ref{tab:overview}.  
We use three sets of configurations generated at bare gauge
couplings $\beta=1.8$, 1.95, and 2.1, corresponding to the lattice
spacings $a\approx 0.22$, 0.16 and 0.11~fm, with lattice dimensions
$L^3\times T = 12^3\times 24$, $16^3\times 32$ and $24^3\times 48$.
The physical lattice sizes are roughly matched at $La\approx 2.5$~fm. 
At each $\beta$, four hopping parameters $\kappa$ are used.
They correspond to $m_\pi/m_\rho \approx 0.8$, 0.75, 0.7 and 0.6. 
The lengths of runs range from 4000 to 7350 Hybrid Monte Carlo
trajectories, and are listed under the column for $N_{\rm Traj}$ 
in Table~\ref{tab:overview}.

\subsection{Propagator Measurements}
\label{subsec:smearingscheme}


We calculate the single quark loop part of the $\eta^\prime$ meson
propagator $\Gconn(t)$ ($=\Gpi(t)$) and the two quark loop part
$\Gdisc(t)$ both for local and smeared sources.  The sink is always
local.  
We use an exponential-like smearing kernel $K$ 
\bdm
\begin{array}{rl}
K(|\vec n - \vec m|\neq 0)&=Ae^{-B|\vec n - \vec m|}\\
K(0)        &=1
\end{array}
\edm
and parameters \ix{A} and \ix{B} are chosen to be the same as 
in Ref.~\cite{cppacsspectrum}.  Gauge configurations are fixed to the 
Coulomb gauge. 

We try applying source smearing to both, one or neither of the quark
propagators. For the pion, Fig.~\ref{fig:pismearcomp} shows a case
where smearing exactly one quark propagator delivers the earliest
plateau. 
A comparison of \ix{\etapr} propagators for different
smearings can be found in Fig.\ref{fig:etasmearcomp}; again a 
preference for smearing is generally seen, although single smearing is not distinguished from
double smearing as it was in the pion case.
Since this trend holds over $(\beta, \kappa)$, we focus on
the single-smeared-local combination for our analysis. 

\figpismearcomp
\figetasmearcomp


Our implementation of the noisy source method is to use 
\ix{U(1)} random sources, fixing to the Coulomb
gauge.   
A $U(1)$ random number $\exp(i\theta(\vec n,t))$ is prepared 
for each site $(\vec n,t)$ of the lattice, and a smeared source is  
made according to 
\begin{equation}
  \eta(\vec n,t) = \sum_{\vec m} K(\vec n - \vec m) \exp(i\theta(\vec m,t)).
\end{equation}
Combining the quark propagator $q(\vec n,t)$ 
for the smeared source $\eta(\vec n,t)$  with $\exp(-i\theta(\vec n,t))$ 
yields the loop amplitude with a single smearing, 
while combining $q(\vec n,t)$ with $\eta^\dagger(\vec n,t)$ gives 
the doubly smeared amplitude.

Our noise sources are generated only at one value of the (spin, color)
pair at a time; other spin and color components of the source are left
at zero. The aim of this procedure is to decrease fluctuations. A
similar procedure has been developed as the ``spin explicit method''
in \cite{smearonecomponent}.  In order to probe the gauge field evenly, we repeat 12
times, varying the spin-color index which is chosen to be non-zero
over all its 12 possible values, generating a different noisy source
for each.  We repeat this whole process \ix{\Nnoise} times, such that
the total number of inversions for a fixed smearing is
$\Ninv=\Nnoise\times\Nspin\times\Ncolor$.

For the coarsest lattice at $\beta=1.8$ the noisy source measurement
is made with four different values of \ix{\Nnoise}, 3, 5, 8 and 20.  
In Fig.~\ref{fig:noisedependenceII} we show how the error in $\Gdisc(t)$ 
varies as a function of $N_{\rm noise}$, taking $t=2$ as a representative 
time slice. In Fig.~\ref{fig:noisedependence} a similar plot is shown for the
$\eta^\prime$ meson mass obtained by fitting the $\eta^\prime$
propagator to a single hyperbolic cosine
function.  
While the error is generally smaller for larger $N_{\rm noise}$ as
expected, the precision increases only slowly with $N_{\rm noise}$,
particularly for light quark masses. Furthermore these errors themselves 
fluctuate significantly. Overall, we cannot be certain that the 
$N_{\rm noise}$ reduction will benefit our result. Since $N_{\rm noise}$ 
and computer time are linearly related, we choose $N_{\rm
noise}=3$ for measurements in the more costly cases of $\beta=1.95$
and 2.1. At \ix{\beta=1.8}, we use \ix{\Nnoise=8}, since it gives 
the smallest observed error in the chiral limit.


\fignoisedependenceII
\fignoisedependence

In our calculation with the local source \cite{cppacspreliminary}, we
evaluate \ix{\Gdisc} with the volume source method without gauge
fixing~\cite{kuramashietal}.  This measurement is made for every
trajectory. On the other hand, \ix{\Gconn} is measured every
\ix{N_{\rm Skip}} trajectories, so an additional binning of
\ix{\Gdisc} is performed in preparation for constructing \ix{\Getaprime}.

We calculate errors of the propagator at each time separation by the
jackknife method.  In our study of flavor non-singlet hadrons 
\cite{cppacsspectrum}, an
autocorrelation analysis has shown that configurations separated by 50
trajectories are sufficiently decorrelated, so we use bins of 50
trajectories in the jackknife analyses, which translates into a bin
size of 5 (\ix{\beta=1.8}, 1.95) or 10 (\ix{\beta=2.1}) measurements.

Slightly different numbers of configurations have been used for the 
\ix{\etapr} in the smeared case. 
The number of measurements of smeared \ix{\etapr} correlators for each $(\beta,
\kappa)$ pair is given under the column for $N_{\rm Smeared~Meas.}$ in
Table~\ref{tab:overview}.

\section{Analysis and Result}\label{sec:analysis}
\subsection{Fitting of Propagators}
\label{subsec:mass}



To extract the $\eta^\prime$ meson mass, we fit the propagator 
using a single hyperbolic cosine function: 
\begin{equation}
G_{\eta^\prime}(t)=A_{\eta^\prime}
\left[\exp(-m_{\eta^\prime}t)+\exp(-m_{\eta^\prime}(T-t))\right] , 
\label{eq:direct}
\end{equation}
where $T$ is the temporal lattice size. 
Alternatively one may fit the ratio $\Gdisc(t)/\Gpi(t)$ to  
\begin{equation}
\frac{\Gdisc(t)}{G_\pi(t)}=1-B\exp(-\Delta m~t) ,
\label{eq:ratio}
\end{equation}
where $\Delta m=m_{\eta^\prime}-m_\pi$.  The $\eta^\prime$ meson mass
is then obtained by adding to \ix{\Delta m} the pion mass \ix{\mpi}
extracted from a standard fit of form 
\begin{equation}
G_{\pi}(t)=A_\pi\left[\exp\left(-m_{\pi}t\right)+\exp\left(-m_{\pi}(T-t)\right)\right] . 
\end{equation}

\figetaefflotwtw \figratlotwtw

Let us first look at data obtained with the local source with 
the volume source method. 
Figure~\ref{fig:etaefflotwtw} shows the effective mass for 
the $\eta^\prime$ for a typical case of $(\beta, \kappa)=(2.1,0.1374)$ on
a \ix{24^3\times 48} lattice. The $\eta^\prime$ effective mass does
not show a clear plateau. We nonetheless try to fit the propagator for
$3\leq t\leq 6$ for this example, in order to compare with the ratio
method.  Figure~\ref{fig:ratlotwtw} shows the corresponding ratio as a
function of time.  A fit of form (\ref{eq:ratio}) over $t\geq
t_{\rm min}$ yields stable values for $m_{\eta^\prime}$ when one
varies $t_{\rm min}$ over $2\le \tmin \le 3 $. This conceals an
unquantifiable systematic error from the effect of excited states.  In
fact, we obtain $\metapr=$ 0.670(26) from the direct fit, while
0.584(16) from the ratio fit, showing a 14\% discrepancy. 

\figetaeffsmtwtw
\figratsmtwtw

On the other hand, fitting propagators from the smeared source 
leads to more reliable estimate of $\metapr$.
We show in Fig.~\ref{fig:etaeffsmtwtw} the $\etapr$ effective
mass for smeared source for the same simulation parameter 
as that for local source above.  There is an apparent plateau 
starting as early as $\tmin=1$.
Fitting with $\tmin=1$ gives $\metapr=$ 0.528(12). 
Figure~\ref{fig:ratsmtwtw} shows the corresponding ratio  
$\Gdisc(t)/\Gpi(t)$ and the fit, which gives $\metapr=$ 0.515(11).
The difference of the two fits remains within 3\%.

\figetacomp
\figetaeffectivemasscomp

In order to compare the quality of data from the local and smeared sources, 
we overlay two propagators and two effective masses 
in Figs.~\ref{fig:etacomp} and \ref{fig:etaeffectivemasscomp},
respectively. The signal for $\eta^\prime$ in the smeared case has
larger errors than the local case. 
On balance, however, the advantage of having
plateaux in the smeared case, as shown in 
Fig.~\ref{fig:etaeffectivemasscomp}, greatly outweighs the disadvantage
of its larger statistical error, which can at any rate be quantified. 
We therefore concentrate on data from smeared propagators in our full 
analyses.

\figetaeffsmze
\figetaeffsmon
\figetaeffsmtw

\tabfitcomppswithfitrangesres

Effective masses from smeared propagator 
at every pair of \ix{\betakappa} are plotted in
Figs.~\ref{fig:etaeffsmze} ($\beta=1.8$), 
~\ref{fig:etaeffsmon} ($\beta=1.95$) and 
~\ref{fig:etaeffsmtw} ($\beta=2.1$). Consulting these effective masses,
we determine $m_{\eta^\prime}$ from fitting to propagators with 
$t_{\rm min}=1$. Numerical values of $m_{\eta^\prime}$
are listed in Table~\ref{tab:fitcomppswithfitrangesres}. 

\figdirratcomp

For completeness, we carry out ratio fits to smeared
propagators with $t_{\rm min}=1$. Numerical values are also given in
Table~\ref{tab:fitcomppswithfitrangesres}. Comparison of results from
the direct and ratio fits are made in Fig.~\ref{fig:dirratcomp}. 
We find that the difference is contained within \ix{8\%}.
We should be aware that the $\pi$ effective mass does not exhibit a plateau 
as early as $t_{\rm min}=1$ even for the smeared source 
(see, {\it e.g.,} Fig.~\ref{fig:pieffectivemasscomp}).  
While the amount of decrease in the pion effective mass is small for the smeared source, 
the ratio fit involves systematic uncertainties of this origin. 
We therefore do not use ratio results in our final analyses.

\figpieffectivemasscomp

\subsection{\label{sec:chiralextrapolation}
Chiral Extrapolation}

To chirally extrapolate the mass of the \ix{\etapr}, we test two
functional forms. One is the form appropriate to a Nambu-Goldstone
boson, with an extra constant term to reflect the non-zero mass of the
\ix{\etapr} in the chiral limit: 
\begin{equation}
(\ametapr)^2= A (\ampi)^2 + B.
\label{eq:NGB}
\end{equation}
We refer to this form as NGB fit. 
Alternatively, one may take $\eta^\prime$ mass itself 
in a linear form in $(\ampi)^2$, which is
standard for vector mesons (non-NGB. fit): 
\begin{equation}
\ametapr= A (\ampi)^2 + B.
\label{eq:nonNGB}
\end{equation}

\figchiralxze
\figchiralxon
\figchiralxtw

Both fitting forms reproduce our data well with comparable
$\chi^2/d.f.=$ 0.9--1.1 (NGB fit) and  $\chi^2/d.f.=$ 0.4--2.2
(non-NGB fit), as shown in Figs.~\ref{fig:chiralxze},
\ref{fig:chiralxon} and \ref{fig:chiralxtw}. 
We use the NGB fit to determine the central values, while
the non-NGB fit is used for estimation of systematic errors.  

In order to extract $\metapr$ at the physical point and in physical
units, we follow the full spectrum analysis 
of Ref.\cite{cppacsspectrum} and set the degenerate $u$ and $d$ quark 
masses and the lattice scale from the ratio of $\pi$ to $\rho$ mass 
\ix{\mpiovermrho = 0.1757} and the $\rho$ mass \ix{m_\rho = 0.7684~\GeV}. 

\tabchiralcomppsres
The $\etapr$ meson mass at the physical point at each $\beta$ is given in  
Table~\ref{tab:chiralcomppsres}. 
The Non-NGB fit leads to values of $\metapr$ larger than 
the NGB fit. The difference is the largest for the coarsest lattice 
(16\% which is about $2\sigma$) and smaller for the two finer lattices
(8\% or $1.5\sigma$ ). 
This difference yields a systematic error of about 4\% ($0.4\sigma$) 
in the continuum limit, as discussed in detail in 
Sec.~\ref{sec:systematicerror}.

\subsection{Continuum Extrapolation}
\label{sec:continuumextrapolation}

\figcontx

Figure~\ref{fig:contx} shows $\metapr$ as a function of $a$.
We use a linear form for the continuum extrapolation, 
\begin{equation}
\metapr= C + D a
\end{equation}
to estimate the central value in the continuum limit, 
since we employ a tadpole-improved value for the clover coefficient 
in our quark action.  The $\chi^2/d.f.$ of the fit is 4.2.
We also try a constant plus quadratic form, since one may expect 
\ix{\Orda} effects which remain after tadpole improvement are small:
\begin{equation}
\metapr= C + D a^2.
\end{equation}
\noindent
We find $\chi^2/d.f.$ of this fit to be 2.8. 
Finally, since the data hardly change
between the finest two lattice spacings, we try removing the coarsest
point and fitting to a constant. The quadratic and constant fits 
are used to estimate the systematic error.



\tabsysone

\subsection{Systematic Error Estimate and Final Result}
\label{sec:systematicerror}

We now consider the systematic errors based on the two variant forms
for the continuum extrapolation, and the alternative form for the
chiral extrapolation.  
In Table~\ref{tab:sysone} we list the four estimates of masses in
the continuum limit. 
The central value is obtained from the linear continuum extrapolation
of results from the NGB fit. The quadratic continuum extrapolation
gives a lower estimate, and the constant extrapolation a still lower one. 
On the other hand, the non-NGB chiral fit gives a raised estimate. 
We therefore take the difference between the central value and 
the value from the constant continuum extrapolation as the lower part
of systematic error, and the deviation of the non-NGB chiral fit as
the upper part.

Our final result for $\metapr$ in the continuum limit reads
\begin{equation}
\metapr = 0.960(87)^{+0.036}_{-0.248} \mbox{\rm \ GeV}.
\label{eq:etaresult}
\end{equation}

\section{Conclusions}\label{sec:conclusion}

We have made the first calculation of the mass for the 
flavor singlet pseudoscalar meson in the continuum limit.  We find
that, using smearing, it becomes possible to fit the flavor singlet
pseudoscalar meson correlator directly with a hyperbolic cosine ansatz. 
Similar conclusions have been reached in Refs.~\cite{duncanetal,sesam}. 


There is a lower systematic uncertainty of nearly \ix{30\%} 
in our final result.  The systematic error breakdown indicates
control over the continuum extrapolation to be the most
important aim for future simulations, although this control may be
established indirectly by some pattern of reduction in statistical
error. On the other hand, the higher systematic error, 
coming entirely from the chiral extrapolation, is only \ix{4\%}.

Our continuum result is in agreement with the experimental value for
the \ix{\etaprime} mass of \ix{\metapr=0.956~{\rm GeV}}, despite 
the fact that our calculation is carried out for 
a flavor singlet meson composed of $u$ and $d$ quarks and within 
two-flavor approxaimation to QCD.
Thus, calculations are underway to attempt to extract the elements of the
mixing matrix between quark-based states and eigenstates of mass
in the framework of the three-flavor QCD with a partially 
quenched strange quark.  We also aim to follow this analysis with a 
continuum result from ``2+1'' flavor QCD, in which the strange quark is
also treated dynamically,  
with similar or better statistics than the present work.

\begin{acknowledgments}
This work is supported in part by Grants-in-Aid of the Ministry of Education 
(Nos.
11640294, 
12304011, 
12640253, 
12740133, 
13640259, 
13640260, 
13135204, 
14046202, 
14740173  
). V. I. L. is supported by the Japan Society for the Promotion of 
Science (ID No. P01182).
\end{acknowledgments}


\end{document}